\documentclass[12pt]{interact}
\usepackage{setspace}
\usepackage{epstopdf}

\usepackage{natbib}
\bibpunct[, ]{(}{)}{;}{a}{}{,}
\renewcommand\bibfont{\fontsize{10}{12}\selectfont}

\theoremstyle{plain}

\theoremstyle{definition}

\theoremstyle{remark}

\usepackage{systeme}
\usepackage{amsmath}
\usepackage{amsfonts}
\usepackage{amssymb}
\usepackage{graphicx}
\DeclareUnicodeCharacter{202A}{FIX ME!!!!}

\usepackage{epstopdf}
\usepackage{subfigure}
\usepackage[T1]{fontenc}
\usepackage[utf8]{inputenc}

\usepackage[usenames, dvipsnames]{color}

\usepackage{fancyhdr}
\begin{document}
		\title{On the statistics of scaling exponents and the Multiscaling Value at Risk}
	
\author{
	\name{Giuseppe Brandi\thanks{Corresponding author. Email: giuseppe.brandi@kcl.ac.uk}\textsuperscript{a} and T. Di Matteo\textsuperscript{a,b,c}}
	\affil{\textsuperscript{a} Department of Mathematics, King's College London, The Strand, London, UK;\\ \textsuperscript{b} Complexity Science Hub Vienna, Josefstaedter Strasse 39, A 1080 Vienna, Austria;\\ \textsuperscript{c} Centro Ricerche Enrico Fermi, Via Panisperna 89 A, 00184 Rome, Italy.}}
	

	
	\maketitle

	\begin{abstract}
	Scaling and multiscaling financial time series have been widely studied in the literature. The research on this topic is vast and still flourishing. One way to analyze the scaling properties of time series is through the estimation of their scaling exponents, that are recognized as being valuable measures to discriminate between random, persistent, and anti-persistent behaviors in these time series. In the literature, several methods have been proposed to study the multiscaling property. In this paper, we use the generalized Hurst exponent (GHE) tool and we propose a novel statistical procedure based on GHE which we name Relative Normalized and Standardized Generalized Hurst Exponent (RNSGHE). This methodology is used to robustly estimate and test the multiscaling property and, together with a combination of t-tests and F-tests, serves to discriminate between real and spurious scaling estimates. Furthermore, we introduce a new tool to estimate the optimal aggregation time used in our methodology which we name Autocororrelation Segmented Regression. We numerically validate this procedure on simulated time series by using the Multifractal Random Walk (MRW) and we then apply it to real financial data. We present results for times series with and without anomalies and we compute the bias that such anomalies introduces in the measurement of the scaling exponents. We also show how the use of proper scaling and multiscaling can ameliorate the estimation of risk measures such as Value at Risk (VaR). Finally, we propose a methodology based on Monte Carlo simulation, which we name Multiscaling Value at Risk (MSVaR), that takes into account the statistical properties of multiscaling time series. We mainly show that by using this statistical procedure in combination with the robustly estimated multiscaling exponents, the one year forecasted MSVaR mimics the VaR on the annual data for the majority of the stocks analyzed.
	
\end{abstract}

\begin{keywords}
Multiscaling time series, Robust scaling, Long-memory, Value at Risk.

\end{keywords}

\section{Introduction}
Nowadays, scaling and multiscaling are widely accepted as empirical stylized facts in financial time series. Since they provide important information to risk and asset managers, they need to be properly addressed and analyzed. The (multi)scaling property of time series is particularly important in risk management and has been recently employed as a warning tool for financial events \citep{antoniades2020use}. In particular, models that implicitly or explicitly assume independence of asset returns should be tested against long-term dependence alternatives. In fact, if the assumption of independence of price increments is not met, risk measures might be severely biased, especially if the long-range dependence is acting with a different degree across the time series statistical moments. In particular, multiscaling has been adopted as a formalism in two different branches of quantitative Finance, i.e. econophysics and mathematical Finance. The former devoted most of the attention to price and returns series in order to understand the source of multifractality from an empirical and theoretical point of view \citep{mandelbrot1963,mandelbrot1967variation, mantegna_stanley_book,dacorogna_book,mantegna_stanley,scaling_review_tiziana,calvet1,lux1,lux_marchesi,tiziana_dacorogna2,buonocore2020} and has recently identified a new stylized fact which relates (non-linearly) the strength of multiscaling and the dependence between stocks \citep{buonocore2020}. The latter instead, builds on the work of \citep{roughvola} on rough volatility and has been used to construct stochastic models with anti-persistent volatility dynamics \citep{roughvola,roughvola2,roughvola3,roughvola4}. Even if the research question comes from different perspectives, it is important to recognize the relevance that its study has in Finance.\par
Multiscaling has been understood to originate from one or more phenomenon related to trading dynamics.\footnote{We refer as trading dynamics the results of the set of actions undertaken by investors in buying and selling financial instruments.} In particular, it can be attributed to the fat tails, the autocorrelation of the absolute value of log-returns, liquidity dynamics, or (non-linear) correlation between high and low returns generated by the different time horizons at which traders operate and the consequent volumes traded. It can also be caused by the endogeneity of markets for which a given order generates many other orders. This occurs especially in markets where algorithmic trading is frequently adopted. There are different methodologies used to compute scaling exponents from time series \citep{sornette2018multifractal}. Among all, let us recall the Multifractal Detrended Fluctuation Analysis (MFDFA) proposed in \citep{kantelhardt}, the Wavelet Transform Modulus Maxima (WTMM) introduced by \citep{muzy1991wavelets,muzy1993}, the Deep learning approach proposed in \citep{corbetta2019deep} and the Structure function approach also known as the Generalized Hurst exponent (GHE) method \citep{van1970structure, kolmogorov1962refinement, tiziana_dacorogna,scaling_review_tiziana}. In a recent paper, \citep{barunik_tails} tested different methodologies against some data specification and empirically showed that the GHE approach outperforms the other models. For this reason, throughout this work, we will use the GHE approach. Notwithstanding the importance of the correct estimation of the Hurst exponent, the analysis has been rarely addressed from a statistical point of view. \par
In this paper, we propose a step-by-step procedure that provides a robust estimation and that tests the multiscaling property in a statistically significant way. Application to simulated data and empirical data allows us also to demonstrate the impact of bias on these estimations. We show how the use of proper scaling and multiscaling can ameliorate the estimation of risk measures such as Value at Risk(VaR). We also propose a methodology based on Monte Carlo simulation, which we name Multiscaling VaR, which takes into account the statistical properties of multiscaling time series by using a multiscaling consistent data generating process. \par
The paper is structured as follows. Sections \ref{sec_ms} and \ref{sec_sp} provide a brief description of multiscaling in Finance and of the statistical procedure proposed to consistently estimate and test the scaling spectrum. Section \ref{sec_se} shows the results of this methodology applied to synthetic data while Section \ref{sec_ea} reports the results of an empirical application to real financial time series. Section \ref{sec_var} is devoted to a practical application of scaling and multiscaling property to VaR while Section \ref{sec_c} concludes. 
\section{Multiscaling in Finance} \label{sec_ms}
In this section, we explain the importance of the multifractal (multiscaling) formalism in financial markets. Let us first fix the notation by defining the prices time series as $P_t$ and the log-prices $p_t=\ln(P_t)$. From this, the log-returns over a time aggregation $\tau$ are $r_{\tau}(t)=p_{(t+\tau)}-p_t$, where $\tau$ is expressed in days. Financial models are usually based on the assumption that log-prices follow a Brownian Motion and the for this model, the rescaled second moment of the log-returns over time aggregation $\tau$ follows
\begin{equation}\label{sqrt}
\sigma_\tau=\mathbb{E}[|r_{\tau}(t)|^2]^{\frac{1}{2}} \sim \sigma\tau^{\frac{1}{2}},
\end{equation}
where $\sigma_{\tau}$ is the standard deviation at aggregation horizon $\tau$ while $\sigma$ is the standard deviation at daily aggregation. This equation is usually referred to as the \textit{square root of time rule} and it is widely applied in quantitative Finance \citep{danielsson2006,wang2011}. Examples are the Black and Scholes model in which the volatility evolves as $\sigma\tau^{\frac{1}{2}}$, or the VaR which under Basel regulatory framework can be computed for higher time aggregation, e.g. the $\tau$ days VaR can be computed as the daily VaR multiplied by $\tau^{\frac{1}{2}}$. In the analysis of the Nile river, Hurst found that the scaling behaviour described by a Brownian Motion was not in line with the empirical data \citep{hurst1956methods}. Scaling and multiscaling analyses have been later introduced in Finance \citep{mandelbrot1963,mandelbrot1967variation,mandelbrot_book,sornette2018multifractal}. To detect multiscaling, it is necessary to study the non-linearity of the scaling exponents of the $q$-order moments of the absolute value of log-returns \citep{mandelbrot1,calvet3,scaling_review_tiziana}. 
In particular, for a process $(p_t)$ with stationary increments, the GHE methodology considers a function of increments \citep{scaling_review_tiziana} of the form 
 \begin{equation}\label{mult_def}
\Xi(\tau,q)= \mathbb{E}\left[|r_{\tau}(t)|^q\right]\sim K_q\tau^{qH_q},
 \end{equation}
 where $q=\{q_1,q_2,\dots,q_M\}$ is the set of evaluated moments, $\tau=\{\tau_1,\tau_2,\dots,\tau_N\}$ is the set of lags used to compute the log-returns, $N$ and $M$ are the maximum numbers of moments and lag specification, i.e. $q_1=q_{min}$, $q_M=q_{max}$, $\tau_1=\tau_{min}$ and $\tau_N=\tau_{max}$, $K_q$ is the $q$-moment for $\tau=1$, and $H_q$ is the so called generalized Hurst exponent which is a function of $q$. Finally, the function $qH_q$ is concave \citep{mandelbrot1,calvet3} and codifies the scaling exponents of the process. A process is uniscaling when the function $H_q$ does not depend on $q$, i.e. $H_q=H$ \citep{scaling_review_tiziana}, while it is multiscaling otherwise. If $H\neq0.5$, the process does not behave as a standard Brownian Motion (Wiener process) and neglecting this feature, would significantly bias the estimation of the true risk. In particular, if $H<0.5$ $(H>0.5)$ the process is said to be anti-persistent (persistent) while if  $H=0.5$ the process can be of two types, i.e. it can have independent increments or it can be a short-term dependent process \citep{lillo2004long,krivstoufek2010long}. Given Equation \ref{mult_def}, a possible way to define a multiscaling proxy is by quantifying the degree of non-linearity of the function $qH_q$. The standard procedure used in order to extract $qH_q$ consists in running a linear regression in log-log scale of Equation \ref{mult_def}, which reads as
 \begin{equation}\label{log_mult_def}
 \ln(\Xi(\tau,q))=qH_q\ln(\tau) + \ln\left(K_q\right),
 \end{equation}
 where $\tau$ is defined in the range $\tau=[\tau_{min}, \tau_{max}]$ and  $q=[q_{min}, q_{max}]$ \citep{scaling_review_tiziana}. A multiscaling proxy can be obtained by fitting the measured scaling exponent with a second degree polynomial \citep{buonocore,buonocore2020} of the form\footnote{Technical details of the choice of this functional form can be found in \citep{buonocore,buonocore2020}.}
 \begin{equation}\label{mult_proxy2}
 qH_q=Aq+Bq^2,
 \end{equation}
 where $A$ and $B$ are two constants. In this mathematical setting, as for different multifractal models in Finance \citep{bacry1,calvet1,calvet4,sornette2018multifractal}, we implicitly assume a quadratic function of $qH_q$. The measured $B$, $\widehat B$, represents the curvature of $qH_q$. If $\widehat B=0$, the process is uniscaling, while if $\widehat B\neq0$, the process is multiscaling \citep{buonocore, buonocore2020}. In order to widely apply the multiscaling formalism in Finance, it is of vital importance the ability to correctly estimate the value of $qH_q$ and consequently, of $A$ and $B$. 
 
 \section{Methodology}\label{sec_sp}
 As highlighted in the previous section, estimating the Hurst exponent from empirical data is a challenging task and these challenges can be categorized in two different classes:
 
 \begin{itemize}
 	\item Those due to the statistical procedure adopted;
 	\item Those linked to the financial data themselves.
 \end{itemize}
 
 Within the first class, we identify two main issues related to the following two points:
  \begin{itemize}
 	\item The statistical model used to compute the scaling exponents;
 	\item The input variables used in the statistical procedure.  
 \end{itemize}
 Within the second class, issues arise mainly from the following question:
 \begin{itemize}
 	\item If the data contain an anomaly, how is this impacting the estimation of the scaling and multiscaling exponents?
 \end{itemize}

In this work, we address the above challenges. First, we focus on the statistical procedure and the implication on financial time series with and without anomalies. Then, we discuss practical implications related to Finance, with special attention to Value at Risk.

\subsection{Statistical procedure}
Multiscaling properties of financial time series have been understood to come from one or more phenomena related to trading dynamics. From the point of view of the financial microstructure, scaling can be attributed to liquidity dynamics, endogeneity of markets, or any other dynamic existing in the market. In particular, the superimposition of distinct strategies and investment horizons generates long-range dependence with different degrees of strength when evaluated at different order moments and this is precisely the definition of multiscaling. In this section we propose a methodology to estimate the Hurst exponent $qH_q$ and the multiscaling depth (curvature) coefficient $B$ in a robust manner. As specified in Equation \ref{log_mult_def}, the estimation of scaling laws is generally performed through a linear regression in log-log scales. The statistical problem which might arise in this context is that the regression is performed minimizing the squared log-errors instead of the true errors. This procedure might, in case of strong deviation from the assumed statistical model for the errors, severely impact the results. The solution to this problem consists of applying a nonlinear regression to the original (i.e. not transformed) data, comparing the fit of the two specifications to the original data and using the one which performs better. Another issue related to the statistical model is the uncertainty associated to the intercept for the $q$ regressions. In particular, we can exactly compute the value of $K_q$ rather than estimating it, thus eliminating possible errors and bias. We can define the standardized $\Xi(\tau,q)$ as 
 \begin{equation}\label{new_model}
\widetilde{\Xi}(\tau,q)= \frac{\Xi(\tau,q)}{K_q},
 \end{equation}
 based on which, Equation \ref{mult_def} can be rewritten as
  \begin{equation}\label{mult_def2}
 \widetilde{\Xi}(\tau,q)\sim \tau^{qH_q}.
 \end{equation}
 Equation \ref{mult_def2} eliminates the possible bias introduced by the estimation of $K_q$ via regression. To easily exploit and model the multiscaling behavior, we define the q-order normalized moment as  
  \begin{equation}\label{new_model3}
 \dddot{\Xi}(\tau,q)=\widetilde{\Xi}(\tau,q)^{\frac{1}{q}}
 \end{equation}
 which transforms Equation \ref{mult_def2} in
  \begin{equation}\label{mult_def3}
 \dddot{\Xi}(\tau,q)\sim \tau^{H_q}.
 \end{equation}
Within this new formulation, the analysis is much easier since now, all the $q$ regressions have a $0$ intercept and the multiscaling is present only if the regression coefficients $H_q$ differ for distinct values of $q$. In fact, for uniscaling time series all regression lines are overlapping while, for multiscaling time series they diverge. Given the formalism introduced by Equation \ref{mult_def3}, it is easy to check whether a process is multiscaling or not. In addition, we can now rewrite Equation \ref{mult_proxy2} for the normalized and standardized structure function of Equation \ref{mult_def3} as
\begin{equation}\label{mult_proxy3}
H_q=A+Bq.
\end{equation}
Even if mathematically equivalent to Equation \ref{mult_proxy2}, this equation has a statistical advantage. Eliminating the multiplication by $q$ from both sides of the Equation, we reduce the possibility of spurious results in case $q$ is a dominant factor in the multiplication. Indeed, the interpretation is equivalent, i.e. $A$ is the linear scaling index while $B$ is the multiscaling proxy. Finally, let us define the relative structure function between two consecutive moments, namely $q_i$ and $q_j$ ( $q_j>q_i$), as follows

\begin{equation}\label{mult_def4}
\dddot{\Xi}(\tau,q_i,q_j)=\frac{\dddot{\Xi}(\tau,q_j)}{\dddot{\Xi}(\tau,q_i)} \sim \frac{\tau^{H_{q_j}}}{\tau^{H_{q_i}}}=\tau^{H_{q_j}-H_{q_i}}=\tau^{H(q_i,q_j)},
\end{equation}
where $H(q_i,q_j)=H_{q_j}-H_{q_i}$. This formalization has a similar structure as the Extended Self Similarity (ESS) methodology \citep{sornette2018multifractal} since the scaling exponents are computed taking as reference another moment function while it diverges from it as the ESS has a reciprocal effect while Equation \ref{mult_def4} has an incremental effect relative to the reference moment. This approach helps in the statistical analysis since we can now test if a process is statistically multiscaling using a significance test on the estimated $H(q_i,q_j)$. In fact, for uniscaling time series we have that $H_q=H$, which implies that the difference between different order moments is always $0$.\footnote{Excluding the special case $H(0,q_1)$.} On the contrary, for multiscaling time series it should be different from $0$ for all $q$. This reduces to a t-test on the regression coefficients estimated using Equation \ref{mult_def4}. Besides the multi-regression approach, it is possible to perform a multivariate regression by rewriting Equation \ref{mult_def4} as

\begin{equation}\label{mult_def5}
\begin{bmatrix}
\dddot{\Xi}(\tau,0,q_1)\\
\dddot{\Xi}(\tau,q_1,q_2)\\
\vdots\\
\dddot{\Xi}(\tau,q_{M-1},q_M)
\end{bmatrix}
=
\tau^{\begin{bmatrix}H_{(0,q_1)}, \vspace{1pt} H_{(q_1,q_2)}, \cdots, H_{(q_{M-1},q_M)}\end{bmatrix}},
\end{equation}
where $M$ is the maximum number of moments used. This is a multivariate nonlinear regression that can be easily solved via a nonlinear optimization algorithm. Such a methodological approach implies a possible relationship between the q-moments used in the regression. Depending on the model assumptions, one can use Equation \ref{mult_def5} or perform $M$ separate regressions for each exponent. In the first case, it is then possible to use an F-test to test if all the coefficients except for the first one ($H(0,q_1)$) are jointly equal to $0$ against the alternative that some coefficients are different from $0$. This is a less restrictive multiscaling test compared to the multiple t-tests. We call \textit{strongly multiscaling processes} those processes which reject both the null hypothesis for all the t-tests and the null of the F-test. Conversely, we call \textit{weakly multiscaling processes} those processes for which the null hypothesis of all the t-tests is rejected but not the null of the F-test.\footnote{If the null hypothesis for one or more t-tests is not rejected but the F-test rejects the null hypothesis, the process is a non-stable multiscaling process.} This is quite intuitive since if a process is multiscaling, all the relative increments are statistically significant. However, if the process reconstructed with a single exponent is statistically equivalent to the one reconstructed with the full multiscaling spectrum, this means that such multiscaling behavior is weak. As already mentioned, it is recommended to estimate the model both in the log-log scale and in the original coordinate system, and to base the choice of the model on a goodness of fit measure.

\subsubsection{The choice of $q$ and $\tau$}\label{qtau}
\begin{figure}[h!]
	\begin{center}
		
\includegraphics[width=0.8\textwidth,height=0.3\textheight]{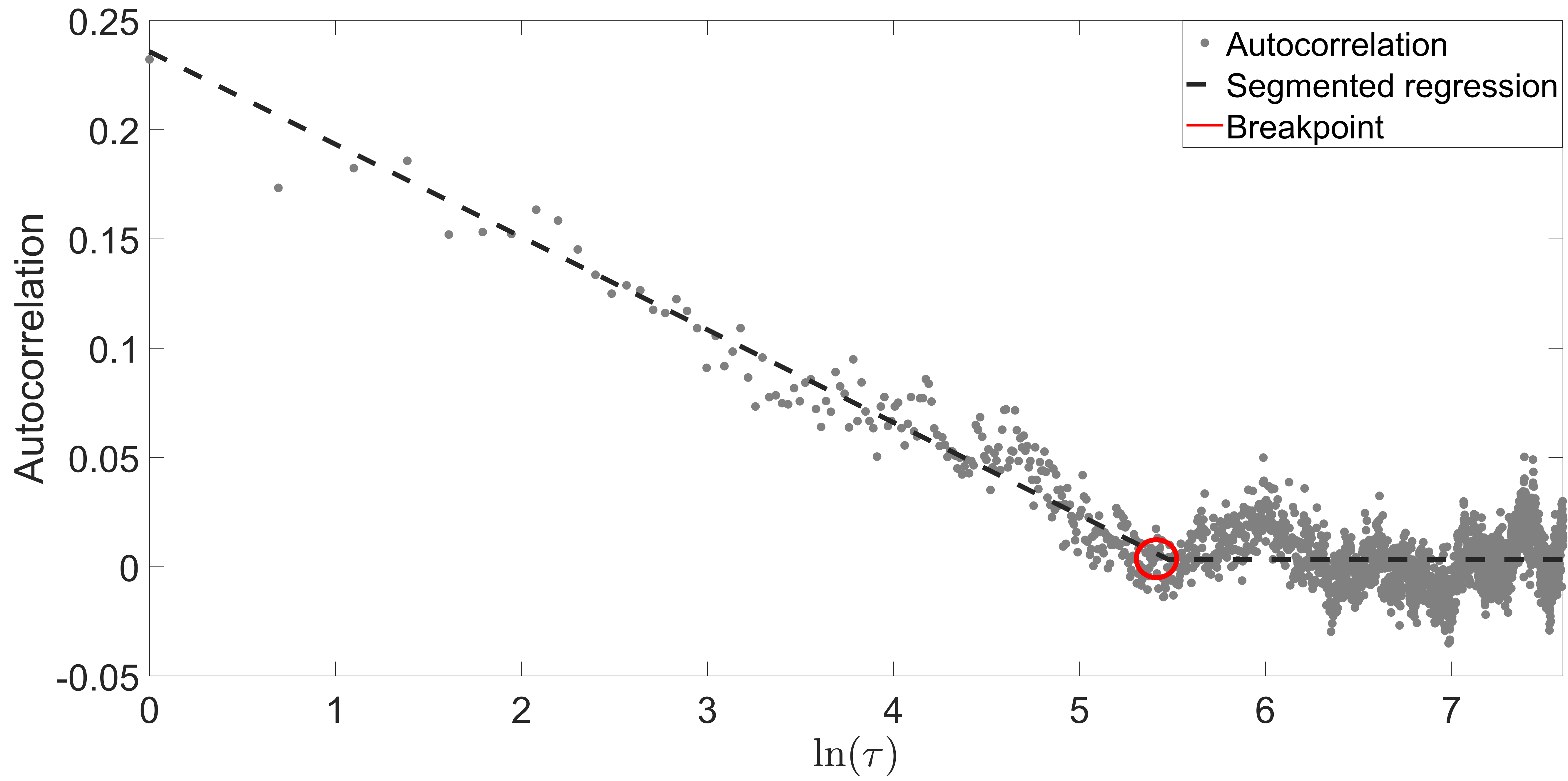}
\caption{ACSR methodology computed on the autocorrelation function of the absolute values of the log-returns time series. The red circle shows the breakpoint where the regression line has a break.}
\label{fig_ACRS2}
\end{center}

\end{figure}

The choice of $q$ and $\tau$ is an important step in the statistical evaluation of the multiscaling exponents. They must be selected using specific statistical criteria. In fact, using the wrong values of $q$ and $\tau$ can severely bias the evaluation. Regarding $q$, many research papers about multiscaling systems propose the use of a vast spectrum of $q$s. This approach has two fallacies. The first one lies in the fact that multiscaling processes are such even for small values of $q$. Secondly and most importantly, given a distribution of returns with tail exponent $\alpha$, for $q\geq\alpha$, $\mathbb{E}[r^q]$ diverges \citep{sornette2018multifractal}. For empirical data, this effect is characterized by a distortion of the moment function which can be misinterpreted as multiscaling even without the presence of temporal dependence. Hence, to have a robust measure of multiscaling, it is necessary to have $q<\alpha$. Any multiscaling behaviour found by neglecting or ignoring this fact, is severely biased and possibly false. The method used to set $q$ can derive from two different approaches, i.e. established research results or direct tail exponent computation. Since it has been empirically shown that financial time series have fat tails with tail exponents ranging from $\sim 1.5$ \citep{scalas1,scalas2,weron} to  $\sim 3$ \citep{sornette2018multifractal}, a conservative approach would be to use $ q\leq 1$. Alternatively, it is possible to estimate $\alpha$ on the empirical distribution through a tail estimator (e.g. \citep{clauset1,clauset2}) and use it as threshold for $q$. In this paper, we follow the conservative approach and use $ q\leq 1$. In fact, if the multiscaling phenomenon is present, it can be extrapolated from this range of moments.\\
Regarding the time aggregation $\tau$, a general rule would be to use the minimum possible value of $\tau$, denoted as $\tau^*$, such that the autocorrelation information of the series is preserved.  The autocorrelation $\rho$ of the return series at lag $\tau$ is defined as:

\begin{equation}
\rho_{\tau}(r(t))=\frac{\mathbb{E}[(r(t+\tau)-\mu)(r(t)-\mu) )]}{\sigma_{\tau}^2}       \end{equation} 	
where $\mu$ and $\sigma_{\tau}^2$ are respectively the mean and variance of $r(t)$. It is a well known stylized fact that returns are expected to be uncorrelated at daily frequency while the absolute and squared returns exhibit long-range persistence \citep{ramacont_review,chakraborti_review}. Among the different procedures used to estimate $\tau^*$, it is worth mentioning:
\begin{itemize}
	\item Segmented regression on the structure function \citep{yue2017linear};
	\item Autocorrelation significance test \citep{buonocore2}.
\end{itemize}
The first procedure computes the structure function for each $q$-moment and succesfully fits a segmented regression in log-log coordinates between $\tau$ and $\Xi(\tau,q)$, and finds two slopes: one for the scaling component and one for the non-scaling component \citep{yue2017linear}. The second approach instead, chooses the value of $\tau$ prior to computing the structure function, setting the value of $\tau^{*}$ as the minimum value of $\tau$ for which the autocorrelation is not statistically significant \citep{buonocore2}. In this paper, we propose a new approach which takes the advantages of  both methods. We name this Autocorrelation Segmented Regression. The rationale behind this approach is to perform a segmented regression on the autocorrelation (or the autocovariance) function computed on the absolute returns and take $\tau^{*}$ as the value which minimizes the sum of squared residuals for the high autocorrelation state and the random noise state, i.e. plateau.\footnote{ \citep{valsamis2019} reviews different segmented regression specifications.} This approach has the advantage of setting the value of $\tau$ in advance, avoiding ad-hock solutions and reducing computations. Nevertheless, the method is less sensitive to a unique non-significant lag. In fact, in noisy data it can happen that for a lag the autocorrelation is not significant while it is significant for a considerable number of subsequent lags. The equation for the proposed Autocorrelation Segmented Regression (ACSR) takes the form
\begin{equation}
\rho_{\tau} (r(t))= \begin{cases} 
\alpha+\beta\tau, & \text{if $\tau<\tau^{*}$} \\
\alpha+\beta\tau^{*}, & \text{if $\tau \geq \tau^{*}$} 
\end{cases}
\end{equation}
where $\alpha$ is the intercept of the regression and can be fixed to be equal to $\rho_{\tau}$ with $\tau=1$, $\beta$ is a slope parameter for the autocorrelation function, $\tau$ is the lag at which the autocorrelation is computed, and $\tau^{*}$ is the value of aggregation which maximizes the autocorrelation information.\footnote{It is important to notice that the segmented regression in the structure function and the ACSR method yields similar results.} We use $\ln(\tau)$ instead of $\tau$ for a better detection of $\tau^*$. Figure \ref{fig_ACRS2} shows how this method works. We generated a process with known $\tau_{max}=250$ and run the ACSR to empirically estimate $\ln(\tau^*)$. As shown in Figure \ref{fig_ACRS2}, we get a value of $\ln(\tau^*)=5.51$ which corresponds to  $\tau^*=247$.   

\subsubsection{Multiscaling estimation and testing procedure}\label{procedure}

Before turning the attention to the simulation experiment, let us recall the full procedure required to robustly extract the scaling exponents:
\begin{enumerate}
	\item Compute $\tau^{*}$ with the Autocorrelation Segmented Regression method;
	\item Compute $q=\alpha$ or rely on the empirical evidence available in the literature;
	\item Perform the linear and nonlinear regressions with the above parameters (Equation \ref{mult_def5});
	\item Asses the goodness of fit of the two models and select the one that overperforms;
	\item Compute the multiscaling curvature using Equation \ref{mult_proxy3} and test for statistical significance.
\end{enumerate}    

Concerning point (5), in this paper we propose a full procedure in order to run what we call the multiscaling test. The testing procedure is divided into four steps. In the first step, we test if each scaling increment $H(q_i,q_j)$ is statistically significant through a t-test. The second step is devoted to the F-test. In particular, we perform the F-test using the predicted relative moments from both the regression with the full estimated scaling spectrum, and the regression where only the first scaling is different from $0$, i.e. $\widehat{\dddot{\Xi}}(\tau,q)$ and $\widehat{\dddot{\Xi}}(\tau,\bar{q})$, where $\bar{q}=\{q_1,0,\dots,0\}$. If the null hypothesis is rejected, that the full spectrum is necessary to recover all the relative moments.\footnote{For the non-linear regression, in order to use the F-test we have to use $\ln(\widehat{\dddot{\Xi}}(\tau,q))$ and $\ln(\widehat{\dddot{\Xi}}(\tau,\bar{q}))$. } The third step of the procedure consists of a random walk (RW) hypothesis. Assuming the multiscaling parameter $B=0$, we perform the regression of Equation \ref{mult_proxy3} with only the constant $A$ and test if $\widehat{A}=0.5$ with a t-test. In fact, for financial returns $H_2=0.5$ which implies $H_2 = 0.5 = A+2B$. Hence, when $B=0$ for monoscaling time series, we expect $A=0.5$. In case the null hypothesis is rejected, this means that the RW scaling is incorrect and the use of the \textit{square-root of time rule} severity creates a bias in the risk measures. The last step involves a confirmatory test of the results deriving from the first and second steps of the test procedure. In particular, we perform the full regression of Equation \ref{mult_proxy3} and test for $\widehat{A}=0.5$ and $\widehat{B}=0$ using a t-test. If this test gives a conflicting result with respect to the first and second steps, we cannot assert anything on the process with precision and a deeper analysis is required by controlling for different input specifications.

\section{Simulation experiment}\label{sec_se}
	In the simulation experiment, we focus on one of the most used models to generate multifractal time series: the Multifractal Random Walk (MRW) proposed by \citep{bacry1,bacry2}. This model is capable to generate multifractal time series with a known multiscaling spectrum. In addition, this model is able to generate time series which are consistent with the financial stylized facts. In the discrete version of the MRW, the process
	$r_{\tau}(t)=p(t+\tau)-p(t)$ is defined as \citep{bacry1}:
 
\begin{equation}
r_{\tau}(t)=p_{(t+\tau)}-p_t=\sum_{k=\frac{t}{\Delta t}+1}^{\frac{t+\tau}{\Delta t}}\epsilon_{\Delta t}(k)e^{\omega_{\Delta t}(k)},
\end{equation}
with
\begin{equation}\nonumber
\begin{array}{cc}
\epsilon_{\Delta t}\sim N(0,\sigma^2\Delta t),	&	 \omega\sim N(-\lambda^2\ln(L/\Delta t),\lambda^2\ln(L/\Delta t))
\end{array}
\end{equation}
where $\lambda$ is called intermittency parameter and determines the strength of the multifractality, $L$ is the autocorrelation length, $\sigma^2$ is the variance of the process, and $\Delta t$ is the discretization step. The distinctive feature of the MRW is that, even if the $\epsilon_{\Delta t}(k)$ are independent, the $\omega_{\Delta t}(k)$ are not, having autocovariance 
\begin{equation}\nonumber
Cov(\omega_{\Delta t}(k_1),\omega_{\Delta t}(k_2))=\lambda^2\ln\rho_{\Delta t}(k_1-k_2),
\end{equation}
whit
\begin{equation}\nonumber
	\rho_{\Delta t}(k_1-k_2)=
\left\{
\begin{array}{cc}
\displaystyle
\frac{L}{(|k_1-k_2|+1)\Delta t} & |k_1-k_2|<L/\Delta t,\\
1 & \mbox{otherwise}.
\end{array}
\right.
\end{equation}
In the continuous limit, the scaling exponents of this model are
\begin{equation}\label{scaling}
\zeta(q)=qH(q)=(\lambda^2+\frac{1}{2})q-\frac{\lambda^2}{2}q^2.
\end{equation}
The power of this model is that it encompasses all the major stylized facts using only three parameters ($\lambda, L, \sigma$). In fact, this model is able to reproduce fat tails, volatility clustering and multiscaling spectrum. 
For the purpose of simulation, we generated $100$ paths each of dimension $T=10000$ and we set the model parameters to $L=250$, $\sigma=1$ and $\lambda=0.05,0.1,0.3$ in accordance to empirical findings \citep{bacry2008log,bacry3,lovsletten2012approximated}. As explained in previous section, $q_{max}=1$. In particular, we use $q \in [0.02,1]$ with steps $0.02$ which converts to $50$ evaluated moments. To select $\tau_{max}$, we use the $\tau^{*}$ estimated by the ACSR. Table \ref{tab_ACSR} shows results for the different specifications of $\lambda$. As shown in the table, the procedure is quite accurate and the 95\% confidence intervals (C.I.) always contain the value of $L=250$ which is the truncation parameter.

	\begin{table}[h!]
		\centering

				\includegraphics[width=0.5\textwidth]{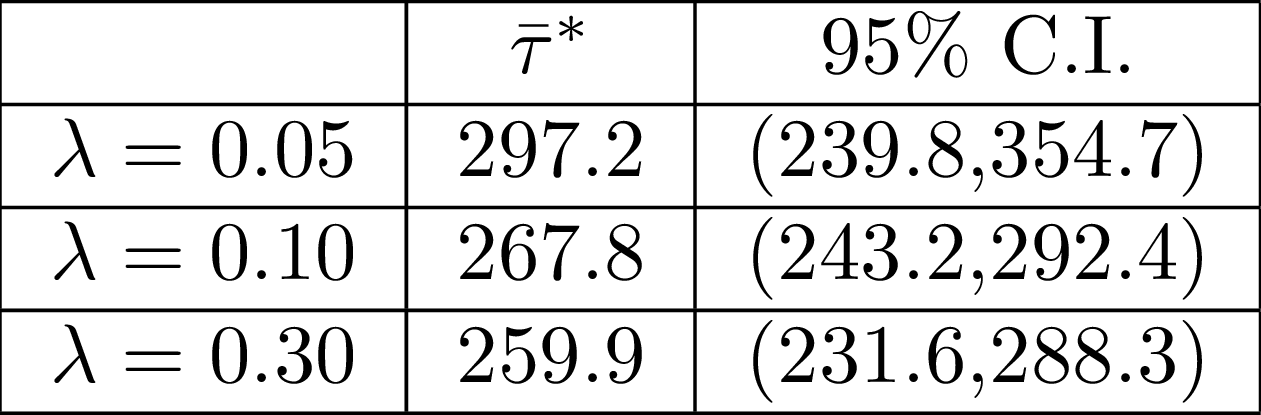}
		
\vspace{5pt}
\caption{Results of the ACSR for the estimation of $\tau_{max}$. $\bar{\tau}^*$ is the mean over all the paths and the $95\%$ C.I. are computed over 200000 bootstrapped samples.}

\label{tab_ACSR}
\end{table} 
Once the parameters are estimated, we compute the multiscaling exponents and evaluate their statistical significance. Since Equation \ref{scaling} gives the true multiscaling spectrum, we can easily test the performance of the GHE approach and compare it with the new proposed methodology of this paper. We use the normalized and standardized structure function (NSSF) proposed in Equation \ref{mult_def3} and the relative normalized and standardized structure function (RNSSF) proposed in Equation \ref{mult_def5}, which we name Normalized and Standardized Generalized Hurst Exponent (NSGHE) and the Relative Normalized and Standardized Generalized Hurst Exponent (RNSGHE), respectively.The latter methodology will be used to test the multiscaling spectrum. Tables \ref{tab_RMSE1} and \ref{tab_RMSE2} present the root mean squared errors (RMSE) of the different methodologies computed over the $100$ realizations for both $A$ and $B$ parameters of Equation \ref{mult_proxy2}.\footnote{For the linear regression case, the NSGHE and RNSGHE are equivalent models, so we report only the result for the RNSGHE.} As it is possible to notice, the RNSGHE generally outperforms with respect to the other specifications. It is important to highlight that by removing the slope ambiguity, results have considerably improved. In fact, the standard GHE approach has the highest RMSE among all the specifications. This result was expected since the new methodology helps to remove uncertainty and thus, to ameliorate the estimation performance. 
\begin{table}[h!]
	\centering
	\includegraphics[width=0.75\textwidth]{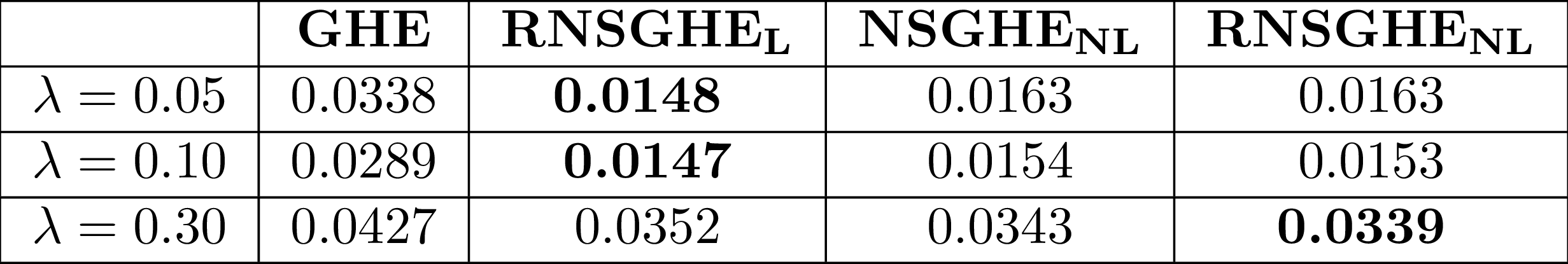}
\vspace{5pt}
\caption{RMSE for the parameter $A$ for the different methodologies. Subscript $L$ refers to the linear regression while $NL$ to the non-linear regression.}

\label{tab_RMSE1}
\end{table}

\begin{table}[h!]
	\begin{center}
	\includegraphics[width=0.75\textwidth]{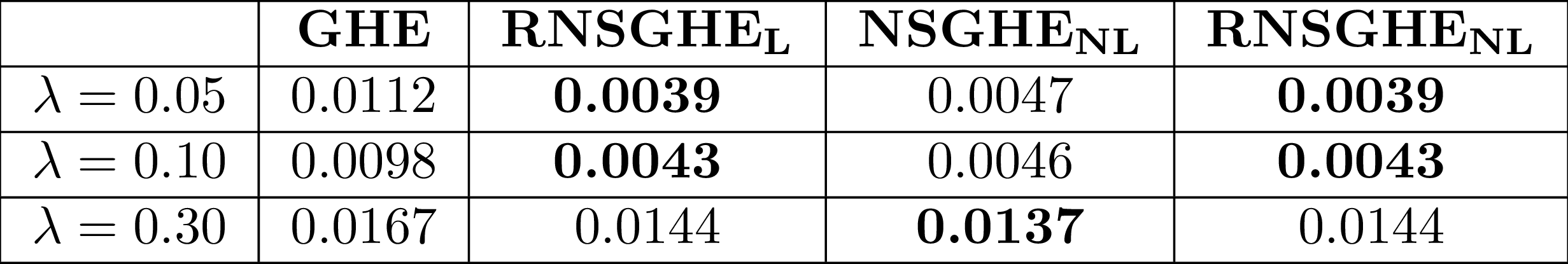}
\vspace{5pt}
\caption{RMSE for the parameter $B$ for the different methodologies. Subscript $L$ refers to the linear regression while $NL$ to the non-linear regression.}
\label{tab_RMSE2}
	\end{center}
\end{table}

\begin{figure}[h!]
	\begin{center}	
		\includegraphics[width=0.8\textwidth,height=0.3\textheight]{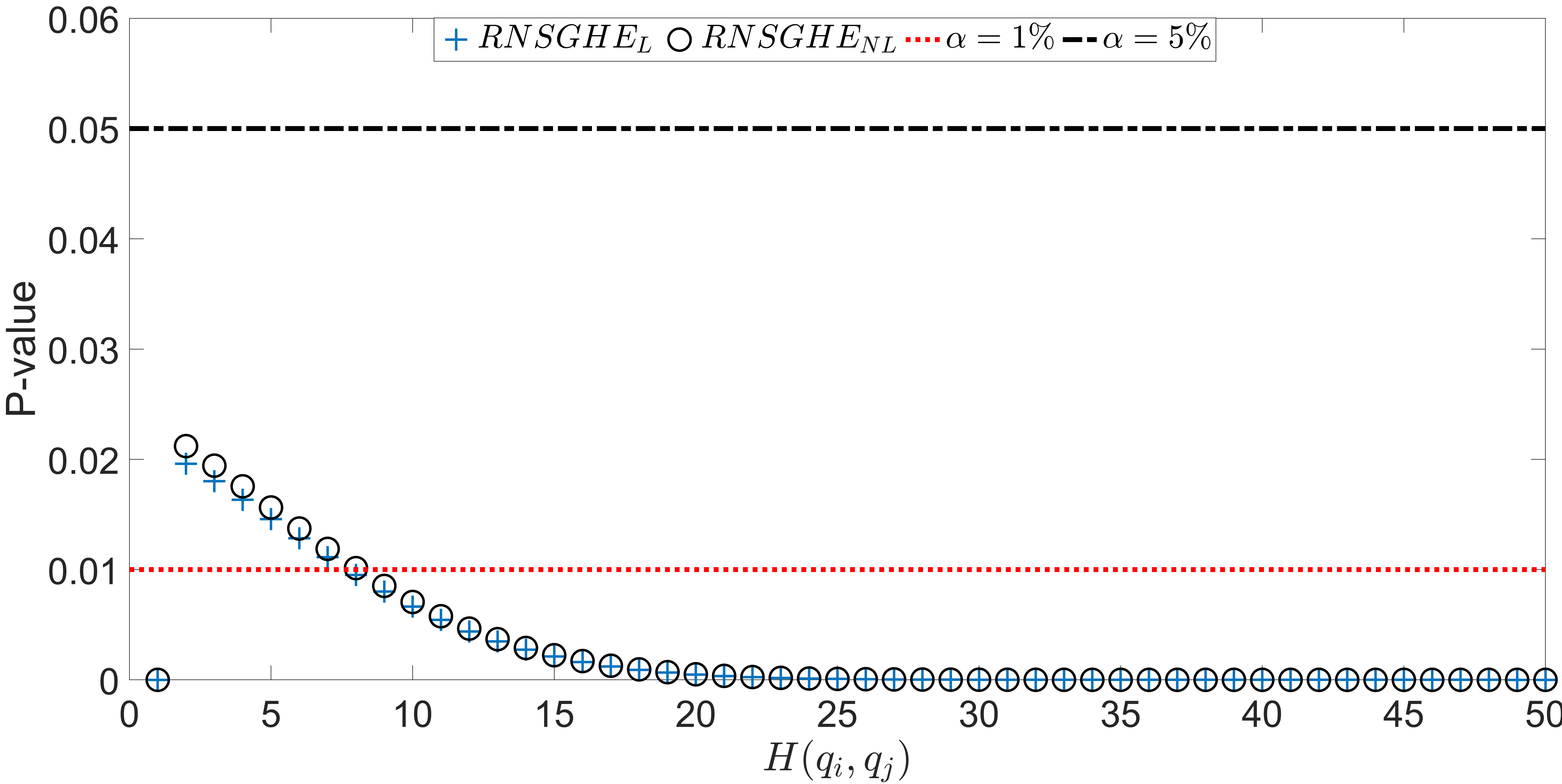} 
	
	\end{center}
	\caption{P-values of all the 50 coefficients related to $q$ moments equations for a multifractal random walk with $T=10000$, $\lambda=0.05$, $L=250$ and $\sigma=1$.}
	\label{ttest_005}
\end{figure}

Tables \ref{tab_RMSE1} and \ref{tab_RMSE2} show a better performance of the RNSGHE\textsubscript{$L$} and RNSGHE\textsubscript{$NL$} compared to the other models, in terms of RMSE. For this reason, we use these models in the paper. Now, we show the nature of the process by performing the multiscaling test. Figure \ref{ttest_005} shows the p-values of all the $50$ coefficients related to the $q$ moments equations for a realization of the MRW model, assuming the same parameters specification as above and $\lambda=0.05$. What we can observe is that by choosing a confidence level of $5\%$, the null hypothesis of scaling increments equal to $0$ for all the evaluated moments is rejected. However, if we set a more stringent confidence level, for example $1\%$, the null hypothesis is not rejected for some coefficients, resulting in a uniscaling process.

For $\lambda=0.1,0.3$ all the p-values are almost $0$. Since we generated processes with a non-negligible amount of multiscaling, this was expected. Once the t-test has been carried out, we perform an F-test on the overall scaling spectrum (Equation \ref{mult_def5}). Results are reported in Table \ref{tabF}. It is possible to infer that the process generated with $\lambda=0.05$ does not reject the null for which all the scaling increments are equal to $0$ while, for $\lambda=0.10, 0.30$ the null is rejected and the full scaling spectrum is necessary to reconstruct the relative moments. By combining this result with the outcome of the t-test, we can conclude that the process generated with $\lambda=0.05$ is weakly multiscaling at $5\%$ confidence level but it is not multiscaling at $1\%$ confidence level. The other two specifications are strongly multiscaling at any reasonable confidence level.
\begin{table}[h!]
	\centering
	\includegraphics[width=0.5\textwidth]{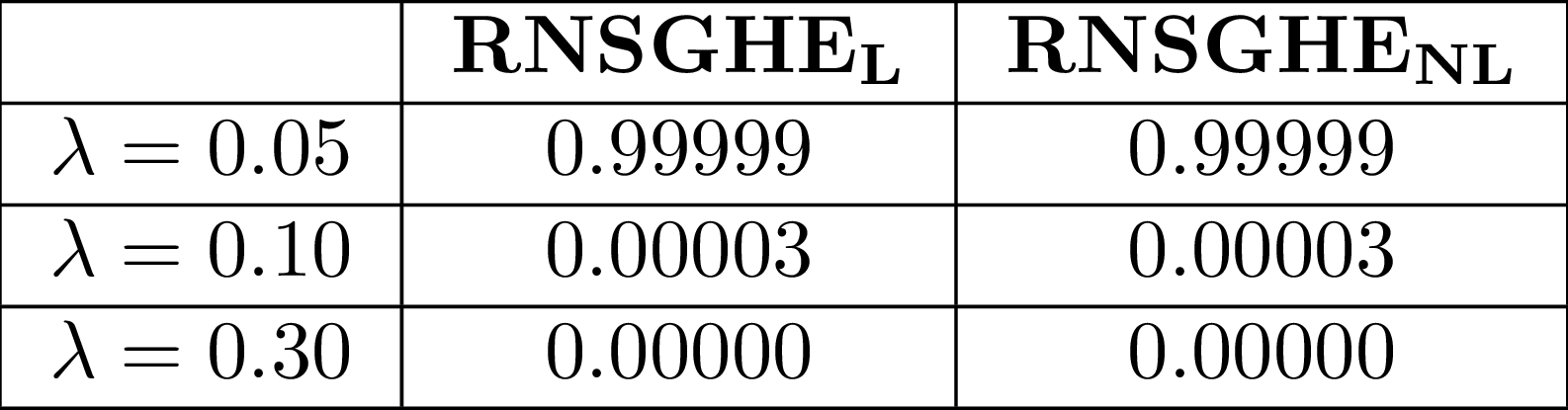}
\vspace{5pt}
\caption{P-values of the F-test for the null that only $H(0,q_1)$ is different from 0. Subscript $L$ refers to linear regression while $NL$ to non-linear regression.}
\label{tabF}
\end{table}

Since these processes have been generated such that they have a specific multiscaling spectrum, the last two tests of the multiscaling test procedure have trivial results.

\section{Empirical application}\label{sec_ea}
In this section, we perform the empirical application of the proposed statistical methodology on financial data described in Section \ref{data}, and produce a statistical analysis of their multiscaling properties in Section \ref{mt}. In Section \ref{anomalies}, we also study how anomalies impact these estimation.

\subsection{Data}\label{data}
The dataset used for the analyses is composed of stocks listed in the \textit{Dow Jones} (DJ). In particular, close prices of stocks are recorded on a daily basis from 03/05/1999 to 20/11/2019, i.e. 5363 trading days. We use 27 over the 30 listed stocks since they are the ones for which the entire time series is available. For the purpose of our analysis, we use log-prices and log-returns. Table \ref{ss} reports the summary statistics of the data.
 
\begin{table}[h!]
	\centering
	\includegraphics[width=1\textwidth]{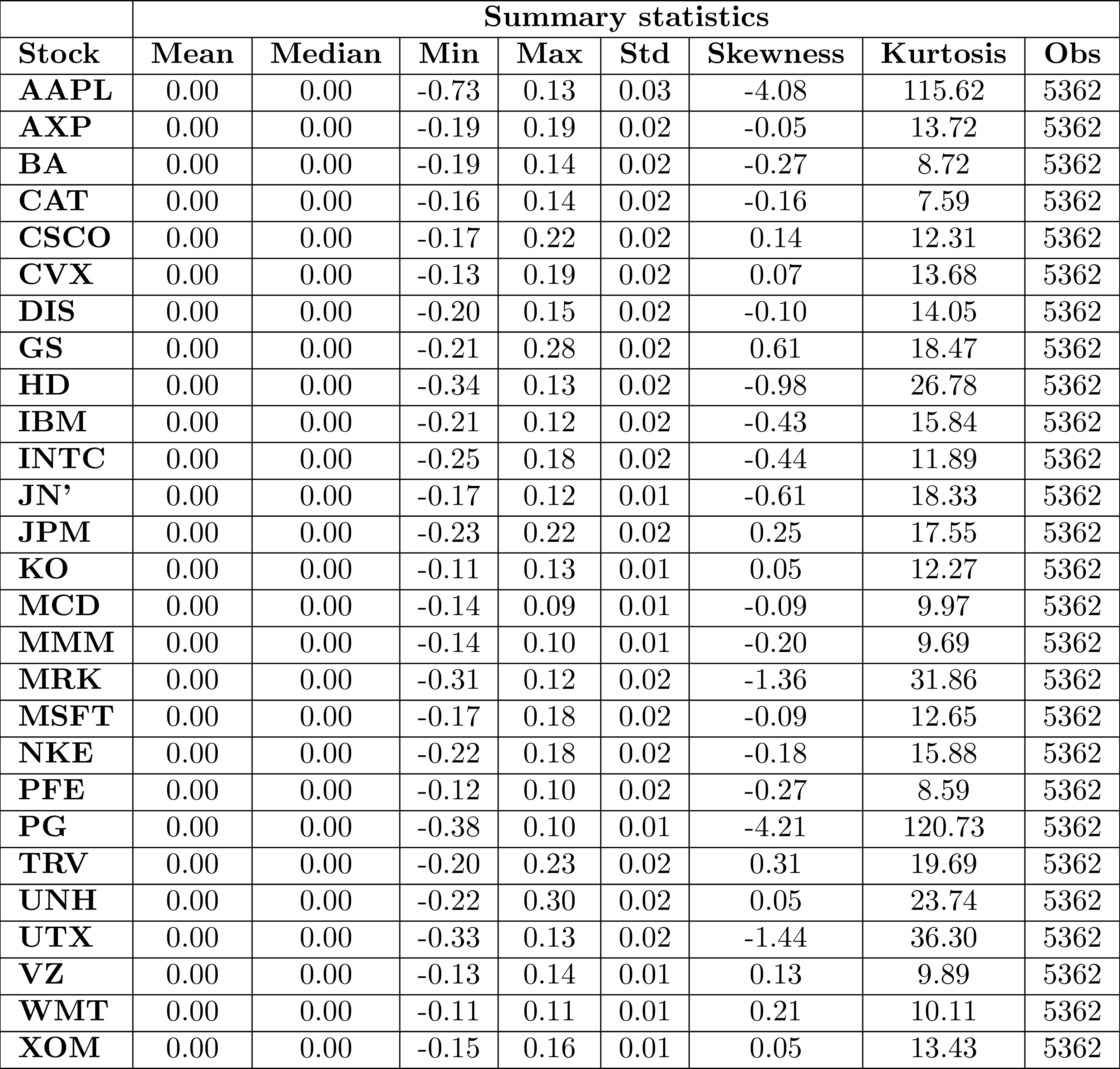}
\vspace{5pt}
\caption{Summary statistics of the log-returns of the analyzed stocks.}
\label{ss}
\end{table}

As shown in this table, all the empirical stylized facts can be observed. Indeed, log-returns are centered at $0$, the Skewness is (in most of the cases) different from $0$, while the high value of the Kurtosis clearly depicts fat tails of the log-returns distributions.

\subsection{Multiscaling test}\label{mt}
In this section, we report results of the multiscaling test.\footnote{All the tests are performed using a confidence level of $5\%$ unless differently stated.} We report the results of all the steps of the testing procedure described at the end of Section \ref{procedure}. Results are presented using the  RNSGHE\textsubscript{$L$} because as explained in the previous section, it has the best performance in the correct estimation of the scaling spectrum.\footnote{It is important to highlight that if the other methods proposed in this paper are adopted, results remain qualitatively unchanged.}
Results are summarized in Table \ref{tab_test}. The second column of the table presents the $\tau^*$ calculated using the ACSR methodology, which is presented in Section \ref{qtau}. For its estimation, we fix a maximum value for the choice of $\tau$ equal to $\frac{T}{5}=1072$ in order not to bias the scaling estimation with too few values. We notice that several stocks reach the boundary value, suggesting a very high rate of persistency in the time series. The third column of this table reports the response to the weak multiscaling (weak M-S) process hypothesis, i.e. hypothesis that a single scaling exponent is enough to approximate the full scaling spectrum but individual scaling increments are statistically significant. As we can observe, none of the analyzed stocks are weakly multiscaling. In fact, as reported in the fourth column, all the stocks pass both the tests and result in strong multiscaling processes. The fifth column of the table reports the result of the RW hypothesis. To perform this analysis we run the regression $H_q=A+\varepsilon$ and test if the estimated $A$,  $\widehat{A}$, is equal to $0.5$.\footnote{This is the standard procedure to estimate the Hurst exponent for uniscaling processes, i.e. $\widehat{A}=\widehat{H}$.} We note that only for two stocks the null hypothesis is not rejected, namely Cisco and Pfizer. However, this is a first order approximation of the process and do not check if the process is multiscaling. The sixth column summarizes results of the confirmation test, which is equivalent to test $\widehat{A}=0.5$ and $\widehat{B}=0$ in the full regression model of Equation \ref{mult_proxy3}. Finally, the last three columns of the table report the estimated Hurst exponent $\widehat{H}$ computed for the RW test, the linear scaling index $\widehat{A}$, and the multiscaling proxy $\widehat{B}$. These results point out that multiscaling is a stylized fact and can be statistically tested by rewriting the structure function in a convenient way.

\begin{table}[h!]
		\includegraphics[width=1.0\textwidth]{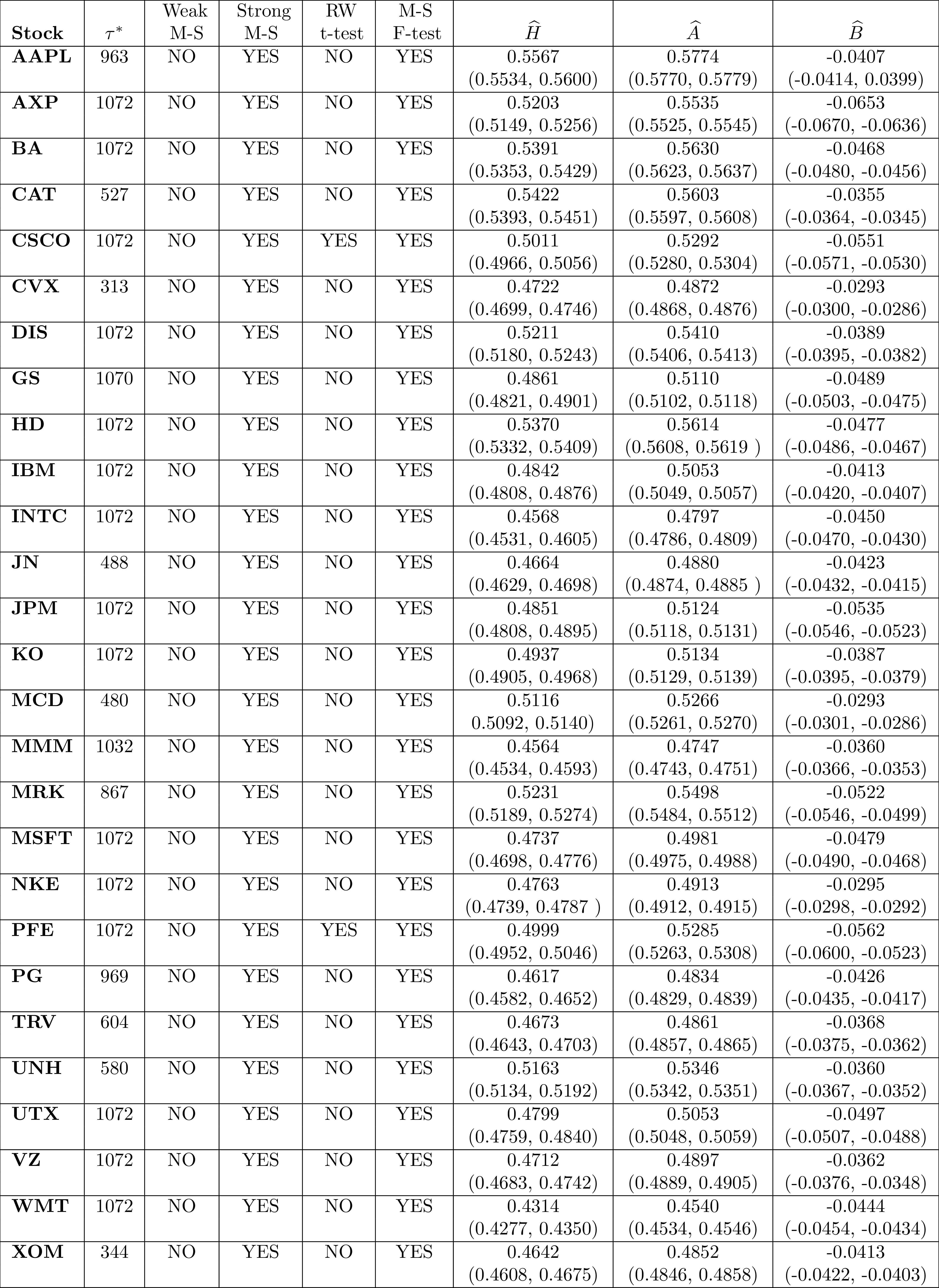}
\vspace{5pt}
\caption{Results of the multiscaling estimation and testing procedure.}
\label{tab_test}
\end{table}

\subsection{Effect of anomalies in the multiscaling estimation}\label{anomalies}
\begin{figure}[h!]
	\begin{center}	
		\includegraphics[width=0.485\textwidth,height=0.2\textheight]{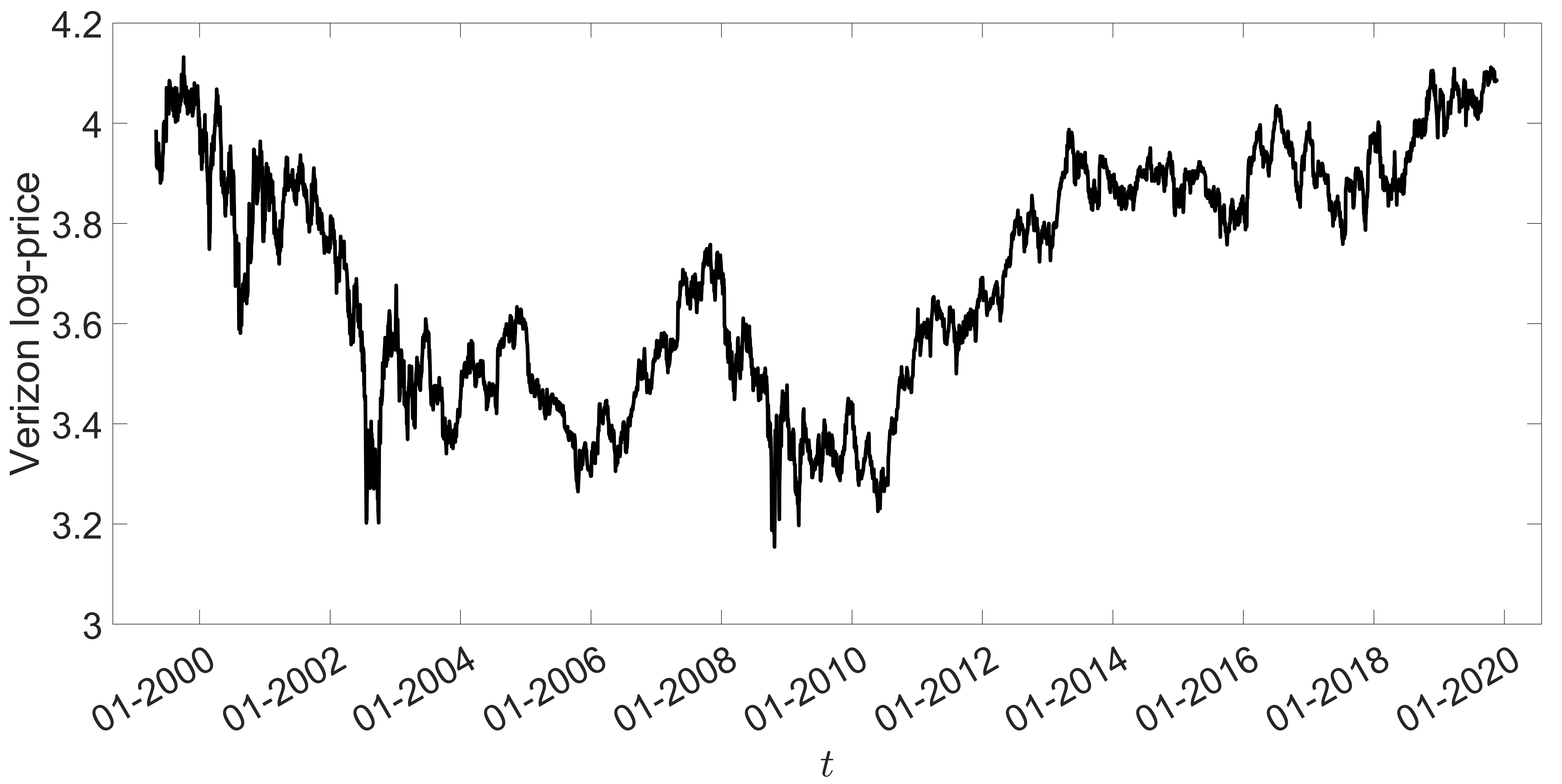} \includegraphics[width=0.485\textwidth,height=0.2\textheight]{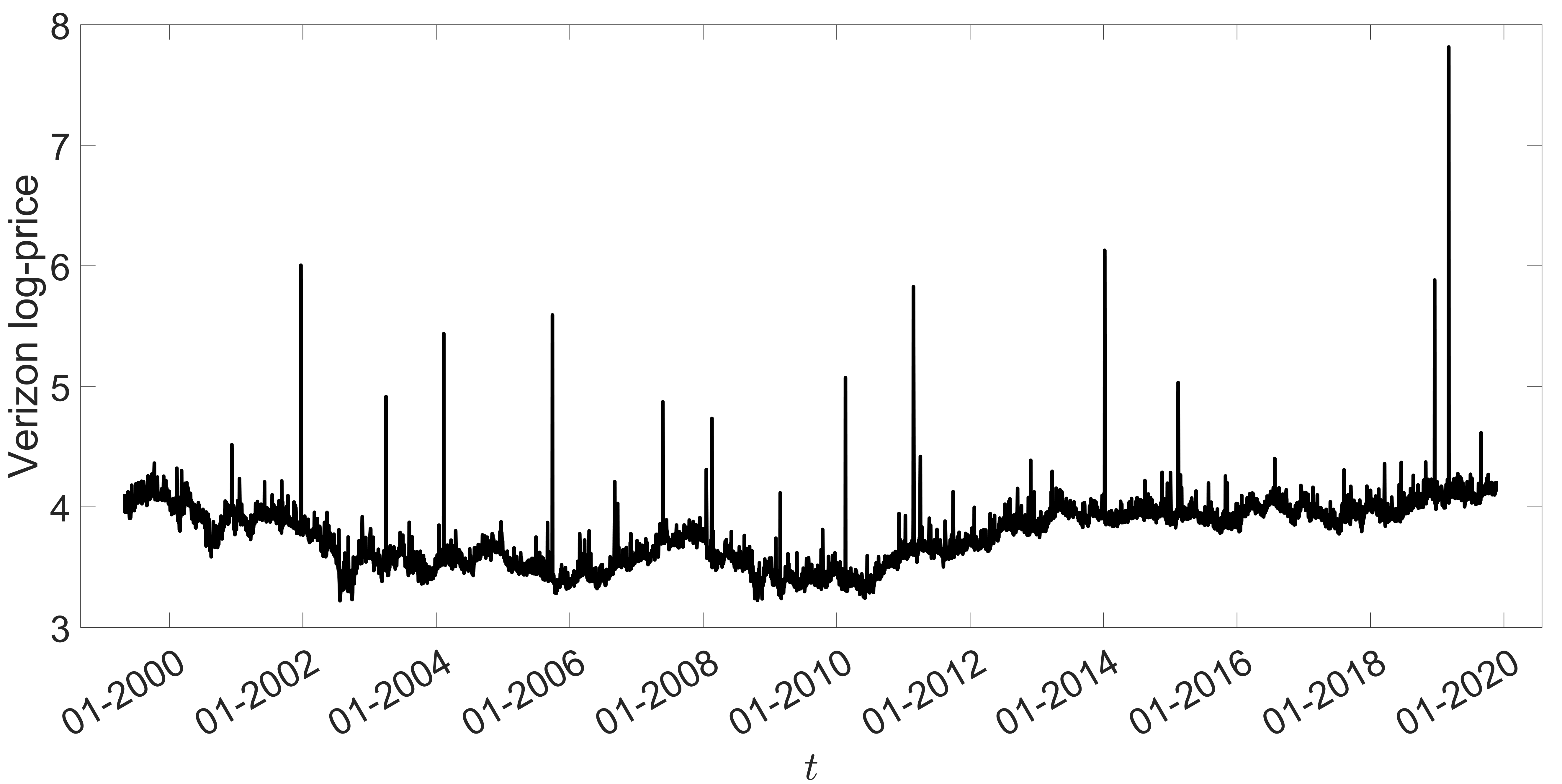}
		
		\includegraphics[width=0.485\textwidth,height=0.2\textheight]{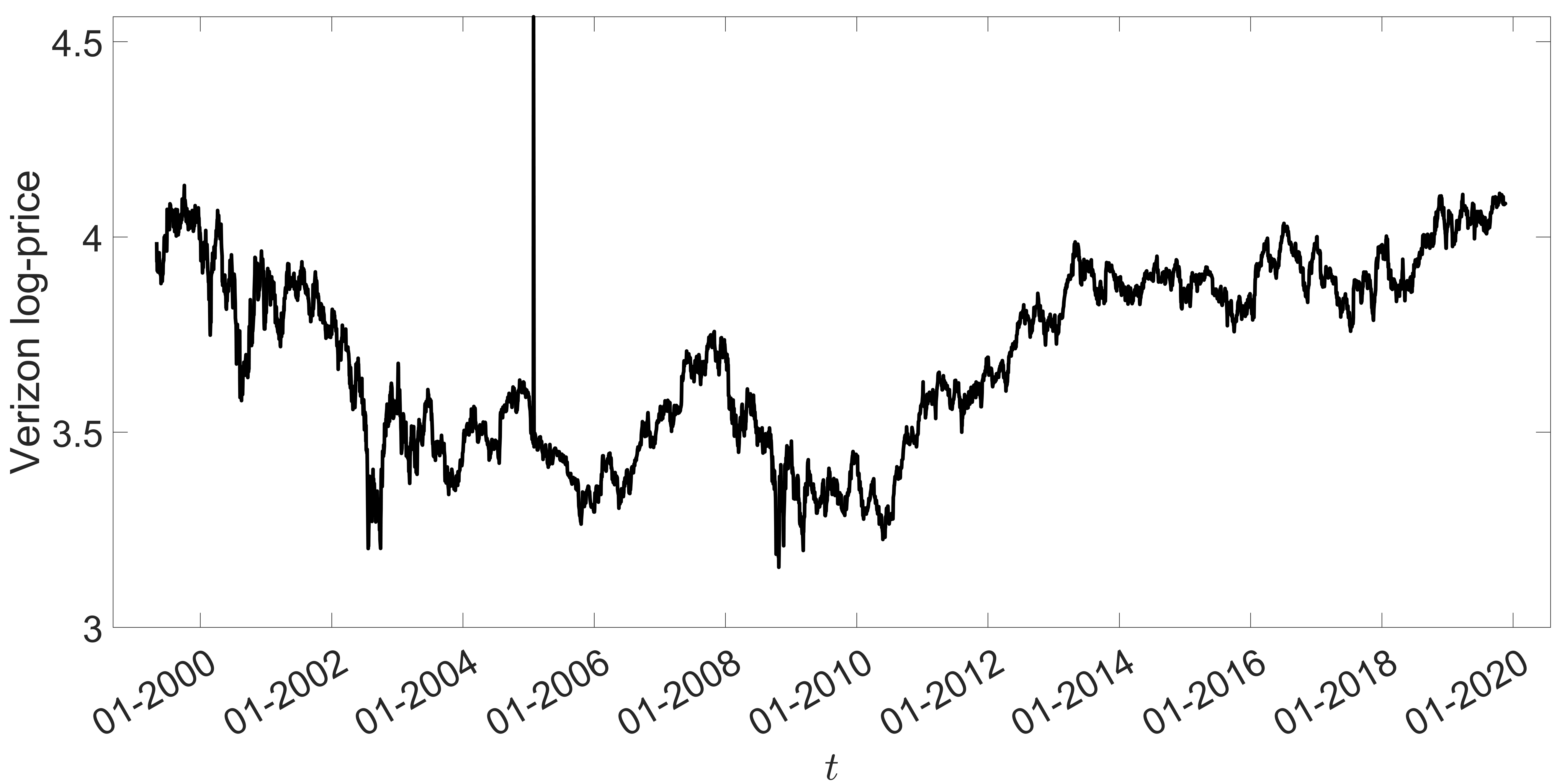}
		\includegraphics[width=0.485\textwidth,height=0.2\textheight]{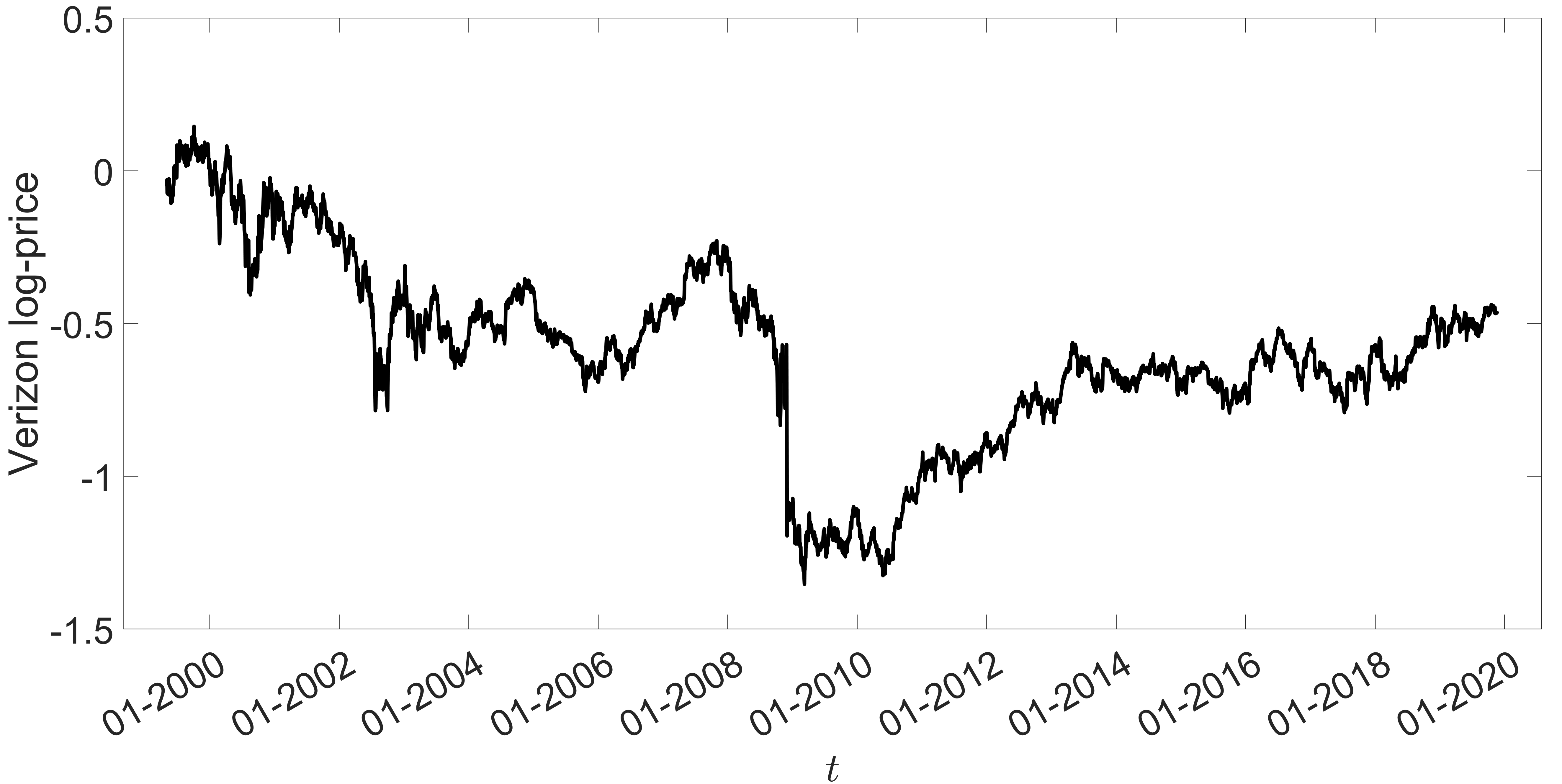}
		\label{fig_jump}
	\end{center}
	
	\caption{Verizon log-prices time series and some anomalies of financial time series.}
\end{figure}

Multiscaling time series are generated from trading dynamics. One of the fundamental aspects of systems exhibiting  multiscaling properties is the strong endogeneity of the sample paths, an aspect which is considered to be originated by financial trading dynamics. For this reason, transient exogenous shocks only distort the analysis and consequently, the estimation procedure. Hence, the statistical procedure used to analyze multiscaling systems are highly sensitive to exogenous shocks. In this context, we refer to an exogenous shock as an unexpected and transient behavior of the stock price, not explainable by the market conditions or by the price path. In addition, anomalies in the time series can occur due to errors or algorithmic trading crashes. Anomalies in financial time series can be grouped in 3 main categories: spikes, jumps and contamination errors. Figure \ref{fig_jump} shows these possible anomalies. The top left panel is dedicated to the original log-price time series for Verizon. This time series is quite volatile and in fact, the log-returns have a Kurtosis index equal to $9$. However, although the distribution of log-returns is fat-tailed, there are not clear anomalies. The top right panel of Figure \ref{fig_jump} depicts the same log-price time series to which a strong fat-tailed series (Kurtosis larger than $1000$) is added. This is the case of contamination error. This is generally due to machine errors in the data transmission process. The bottom left panel reflects the Verizon log-price time series with a random spike added. The spike can arise from multiple sources, among all algorithmic trading errors or contamination errors due to data manipulations. The last panel in the bottom right corner represents the log-price series with an added jump. Jumps per se can arise from endogenous or exogenous shocks. However, if they derive from an endogenous driving force, they persist in the jump direction. Conversely, if they come from an exogenous source instead, they tend to be transient. In a relatively recent paper \citep{muzy2003}, the authors explain that huge financial crashes can be originated from endogenous shocks which have a huge persistence behavior. These kinds of shocks are inherent in the price process so they are not transient anomalies.

\begin{figure}[h!]
	\begin{center}	
		\includegraphics[width=0.8\textwidth,height=0.3\textheight]{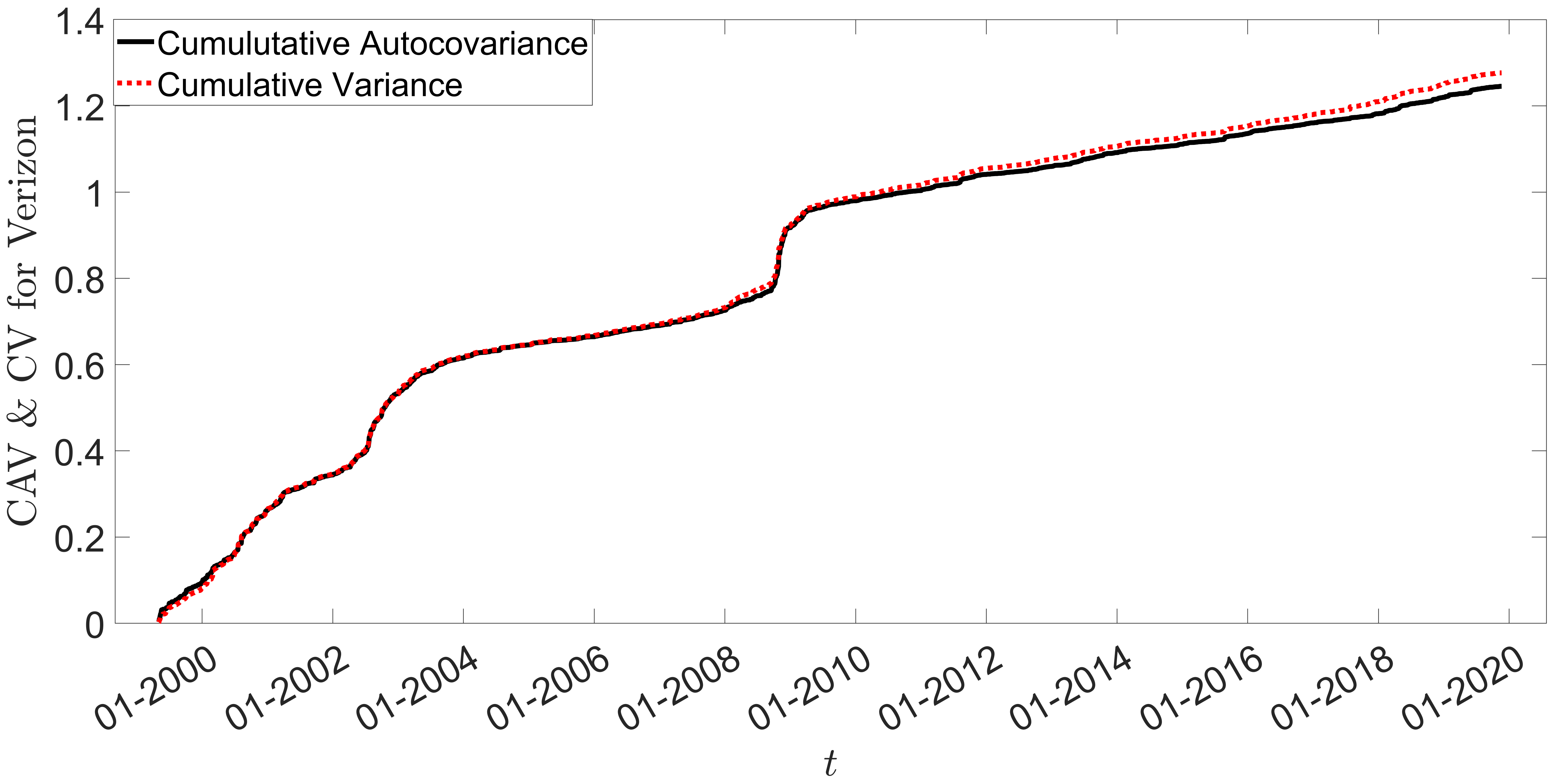} 
	\end{center}
	\caption{Cumulative Variance (CV) and Cumulative Auto-Covariance (CAV) for the Verizon time series.}
	\label{cav}
\end{figure}

In a mostly technical paper, \citep{katsev2003hurst} show both theoretically and experimentally that such data anomalies can strongly bias results, especially for short datasets. In particular, the paper shows that under certain circumstances, these irregularities can generate spurious scaling. For these reasons, it is suggested to analyze the time series and eliminate such anomalies before proceeding with the scaling estimation. In order to do so, we propose a methodology based on financial stylized facts. More precisely, we use volatility clustering and long-range dependence of asset returns \citep{ramacont_review,chakraborti_review}. In this empirical context, the quantities that we name Cumulative Variance (CV) and Cumulative Auto-Covariance (CAV):\footnote{In the financial High-frequency literature, these quantities are strongly related to the Realized Variance and Bipower Variation.}
\begin{equation}
CV(t)=\sum_{i=1}^{t}r_i^2    
\end{equation}
and 

\begin{equation}
CAV(t)=\sum_{i=1}^{t}|r_{i+1}||r_{i}|    
\end{equation} 
should be very similar, except when an exogenous (unexpected) anomaly exists. The volatility clustering drives the similarity in the short period since $|r_{i+1}|$ and $|r_{i}|$ are expected to be very similar (same cluster), while the long-range dependence drives the similarity of the two measures over the long-run. Figure \ref{cav} represents the two quantities for the Verizon stock. These two quantities are approximately equal, confirming that even with high volatility and many tail events, the time series does not contain exogenous shocks. 

\begin{figure}[h!]
	\begin{center}	
		\includegraphics[width=0.8\textwidth,height=0.25\textheight]{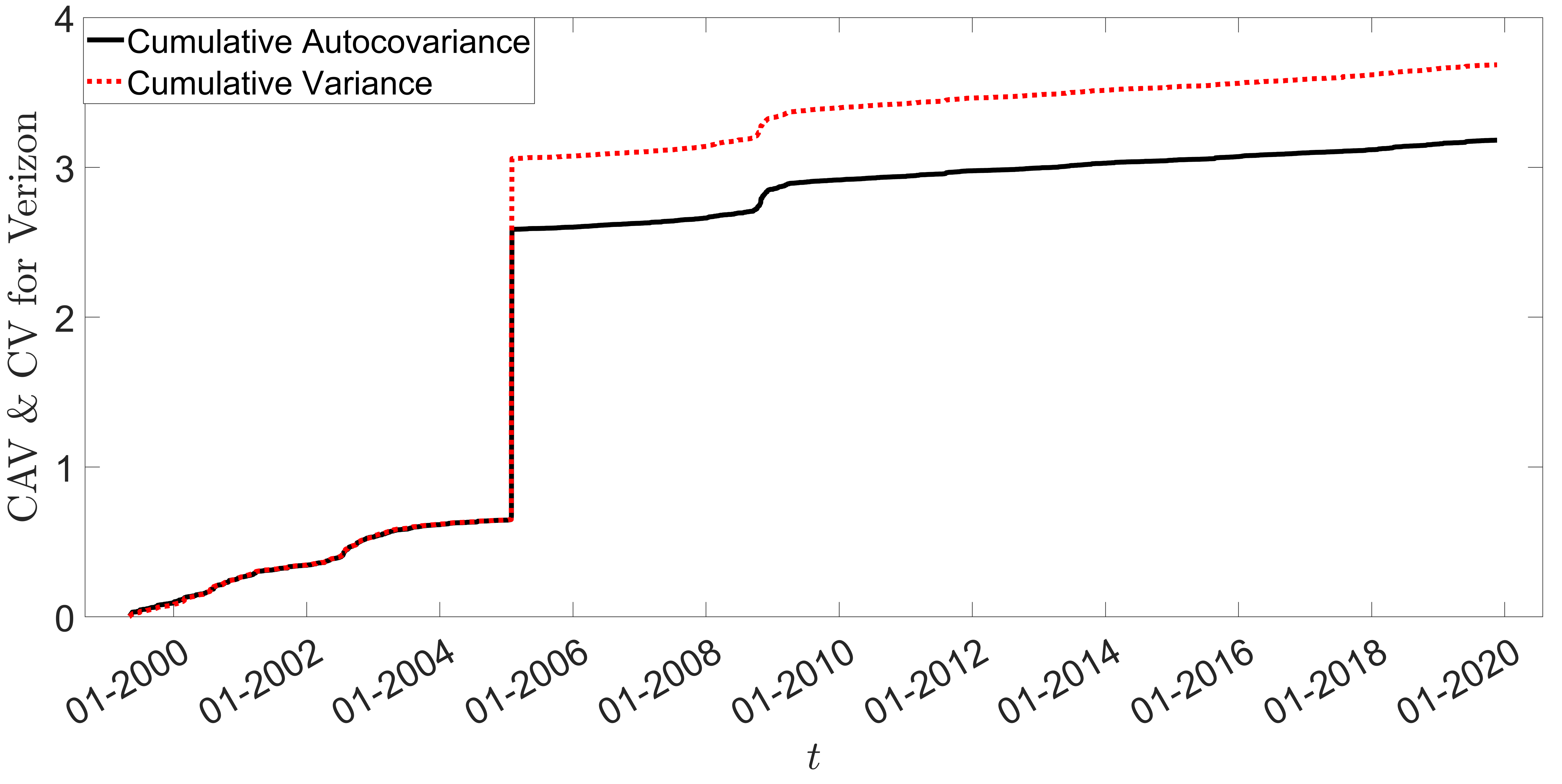} 
		\includegraphics[width=0.8\textwidth,height=0.25\textheight]{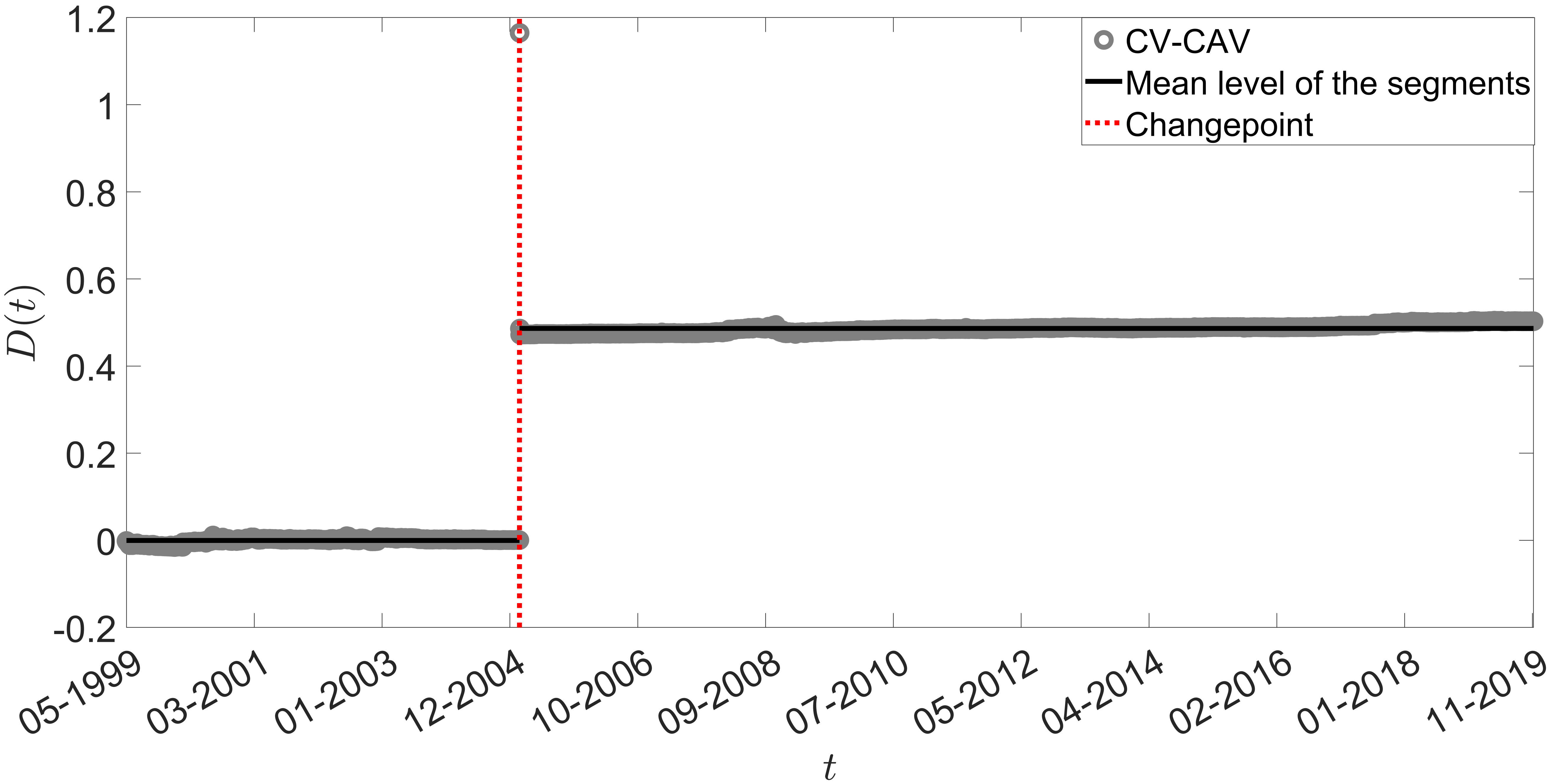}
		
	\end{center}
	\caption{Cumulative Variance ($CV(t)$) and Cumulative Auto-Covariance ($CAV(t)$) for the Verizon time series (top panel) and the measure $D(t)$ (bottom panel) in case of the added spike.}
	\label{Ds}
\end{figure}

\begin{figure}[h!]
	\begin{center}	
		
		\includegraphics[width=0.8\textwidth,height=0.25\textheight]{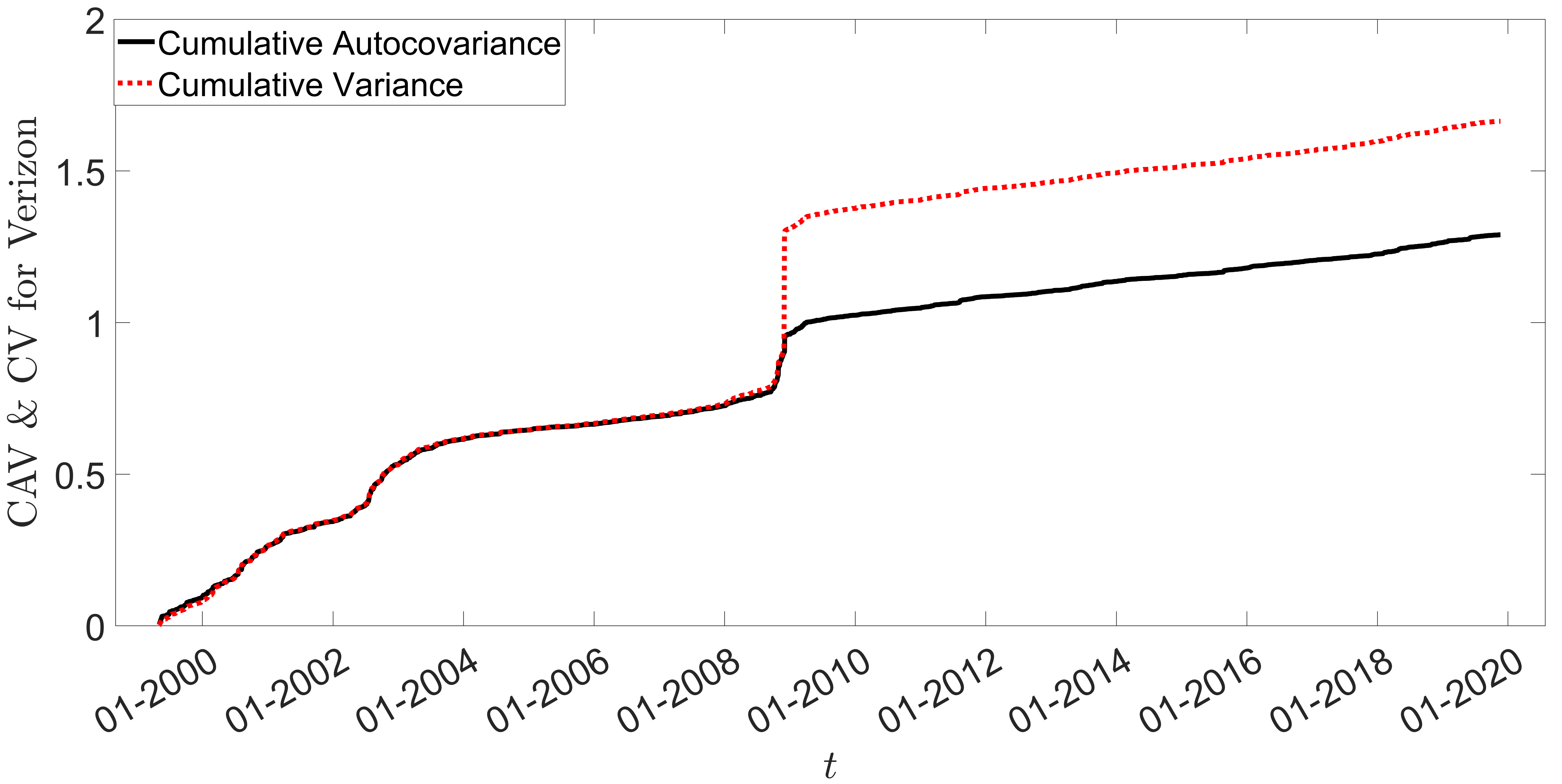} 
		\includegraphics[width=0.8\textwidth,height=0.25\textheight]{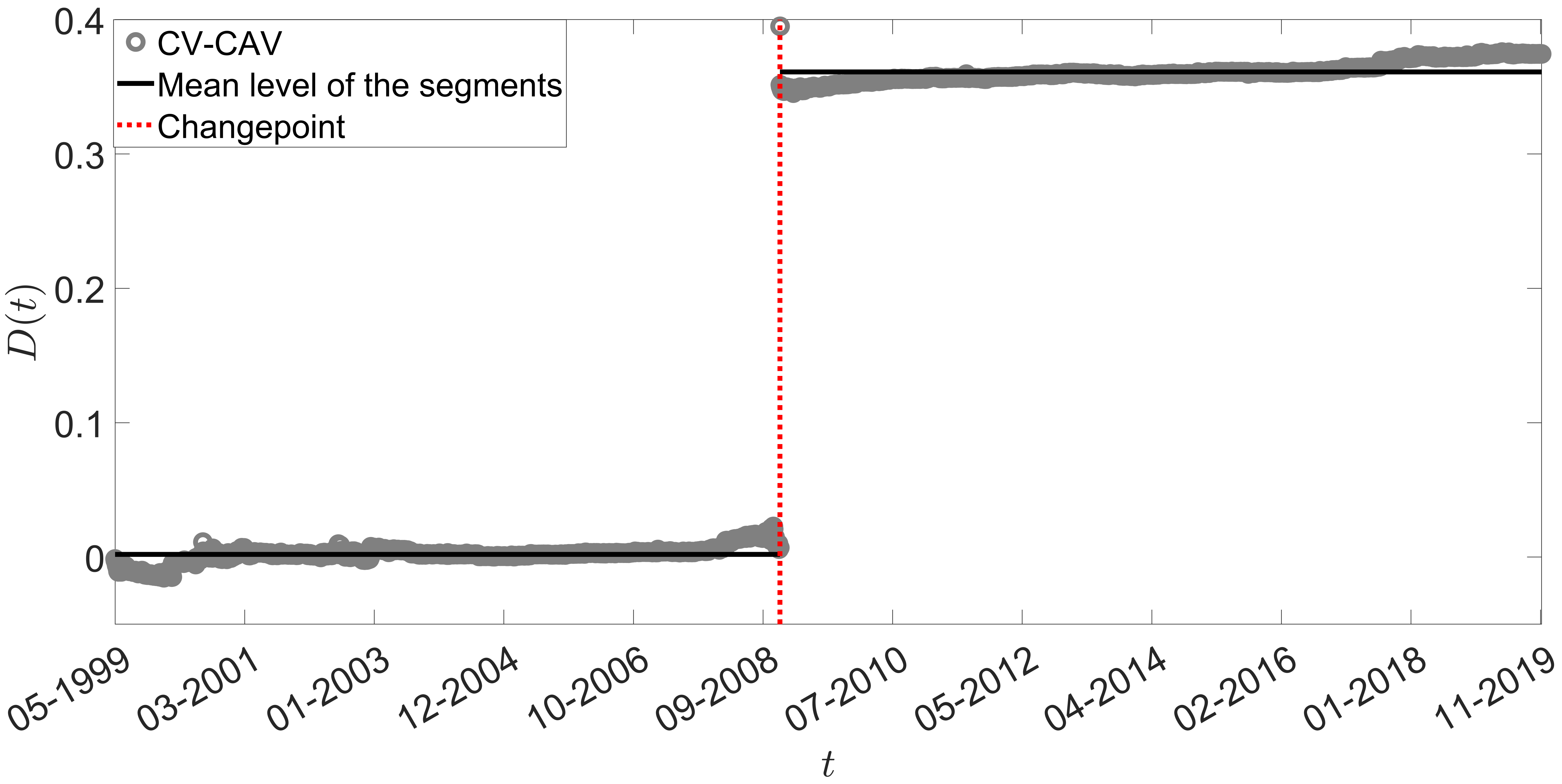}
	\end{center}
	\caption{Cumulative Variance ($CV(t)$) and Cumulative Auto-Covariance ($CAV(t)$) for the Verizon time series (top panel) and the measure $D(t)$ (bottom panel) in case of the added jump.}
	\label{Dj}
\end{figure}

The difference between the two cumulative series is given by $D(t)=CV(t)-CAV(t)$. By running a change-point detection in the intercept and slope of $D(t)$, it is possible to detect the anomalies in the price series and replace the corresponding values on the original series according to a specific rule, e.g. the mean of previous and subsequent data points. Top panels of Figures \ref{Ds} and \ref{Dj} reflect the case in which a spike and a jump have been added to the log-price time series of Verizon, respectively. As shown in the figures, the two measures start to diverge significantly exactly in correspondence of the anomaly point. The bottom panels instead, represent the change point detection performed on the quantity $D(t)$. The panel shows that the procedure correctly identifies the position of the anomaly. 

Given the fact that such events (spikes or jumps) are rare and have unconventional magnitude, their removal can only benefit the analysis. Let us estimate the multiscaling exponent for the Verizon stock when the anomalies reported in Figure \ref{fig_jump} (bottom panels) are not removed by the time series. Table \ref{tab_an} reports the results. The estimated values change considerably, especially in the scenario where a spike is added. 

\begin{table}[h!]
	\centering
	\includegraphics[width=0.5\textwidth]{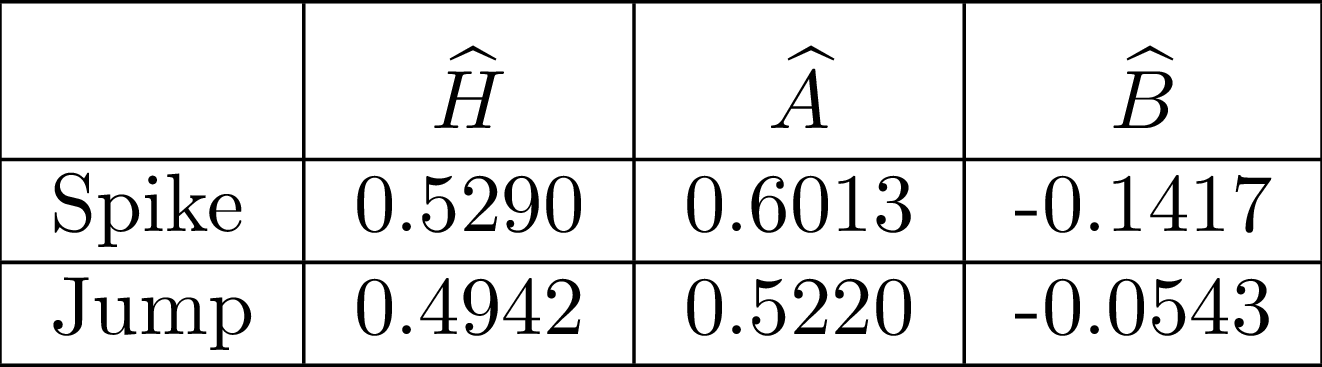}
	\vspace{5pt}
	\caption{Results of the multiscaling estimation on the times series reported in Figure \ref{fig_jump} (bottom panels) with anomalies not removed.}
	\label{tab_an}
\end{table}
For completeness, we also perform a t-test with the null hypothesis of no difference between the estimates with anomalies and the estimates reported in Table \ref{tab_test}. The null hypothesis for all the coefficients is strongly rejected at any confidence level. 

To show how these anomalies can generate spurious multiscaling, we generate $100$ fractional Brownian motions (uniscaling process) of length $1000$ with Hurst exponent $H=0.47$ (the one estimated for Verizon). To these simulated time series we add a spike and a jump and estimate $\widehat{H}$, $\widehat{A}$ and $\widehat{B}$ for both the series with and without the anomalies. Results are reported in Table \ref{table_fbm}. As we can see, when the anomalies are not present in the time series, the average values for $\widehat{H}$, $\widehat{A}$ and $\widehat{B}$ are in line with the true values and not statistically different from them. In the scenario with the added spike, we can see that $\widehat{H}$, $\widehat{A}$ and $\widehat{B}$ are severely biased. In particular, due to the spike, the times series look multiscaling, while it is not. Finally, for the case of the added exogenous jump, we have that the scaling exponent curves and the parameter $B$ is not equal to $0$. Also the other two estimated coefficients are upward biased and statistically different from the theoretical ones.
\begin{table}[h!]
	\centering
	\includegraphics[width=0.8\textwidth]{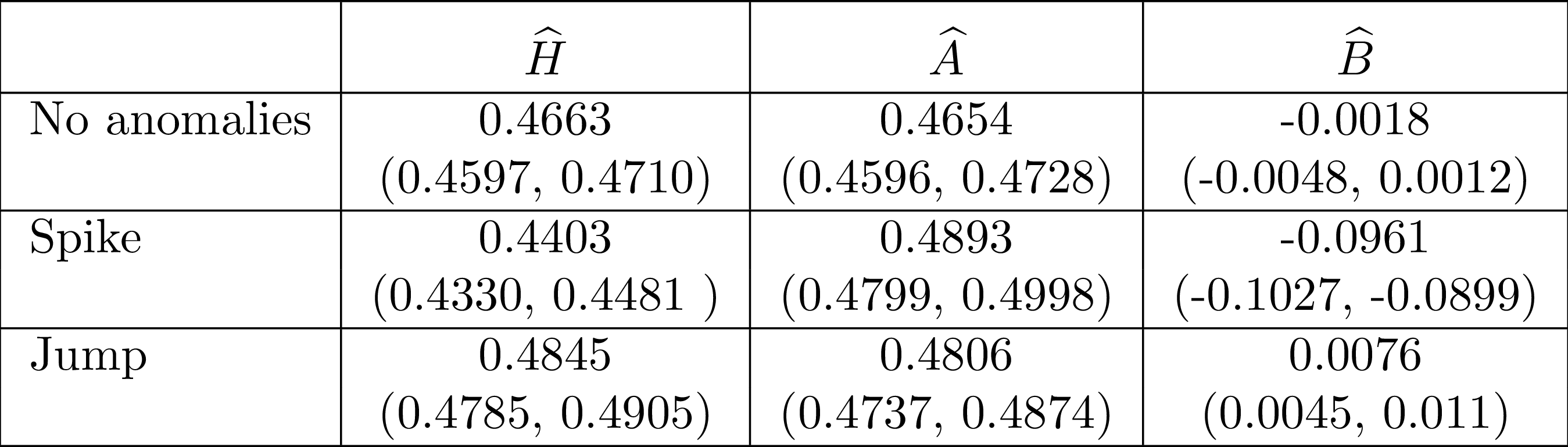}
		\vspace{5pt}
	\caption{Average of the 100 estimated $\widehat{H}$, $\widehat{A}$ and $\widehat{B}$ for a fractional Brownian motion with $H=0.47$. $95\%$ C.I. computed over 200000 bootstrapped samples are reported in parenthesis.}
	\label{table_fbm}
\end{table}
These results clearly show that the scaling exponents are sensitive to such anomalies and estimates can be biased if such anomalies are not carefully analyzed.
\subsection{Jump test on the empirical data}
We performed the jump detection analysis on the 27 stocks used for the analysis and found that 6 of them exhibited jumps. Table \ref{tab:jump_res} summarizes the results. We report the date(s) in which the jump is identified by the change point detection algorithm,\footnote{For the change point detection we used $\frac{1}{2}\parallel D(t)-\mathbb{E}\left[D(t)\right]\parallel_2$ as penalty.} as well as the estimated $\hat{H}$, $\hat{A}$ and $\hat{B}$ when the jump(s) have been removed.\footnote{We imputed the value corresponding to the jump with the average between previous and subsequent data points.}

\begin{table}[h!]
	\centering
	\resizebox{0.7\textwidth}{!}{%
		\begin{tabular}{|l|c|c|c|c|c|}
			\hline
		          & Jump date             & $\widehat{H}$                     & $\widehat{A}$ & $\widehat{B}$ \\ \hline
			AAPL               & 29-Sept-2000                      & 0.5315    & 0.5527    & -0.0416   \\ \hline
			HD                 & 12-Oct-2000                      & 0.5203    & 0.5535    & -0.0653   \\ \hline
			MRK                & 30-Sept-2004                      & 0.5391    & 0.5630    & -0.0468   \\ \hline
			PG                 & 07-Mar-2000                      & 0.5422    & 0.5603    & -0.0355   \\ \hline
			UNH                & 13-Oct-2008					  & 0.5011    & 0.5292    & -0.0551   \\ \hline
			UTX                & 17-Sept-2001 					  & 0.4722    & 0.4872    & -0.0293   \\ \hline
		\end{tabular}%
	}
	\vspace{3pt}
	\caption{Results of the jump detection analysis and estimated exponent parameters}
	\label{tab:jump_res}
\end{table}

From Table \ref{tab:jump_res}, we can notice the following: when removal of the jump increases the estimate of $\widehat{H}$ and $\widehat{A}$, multiscaling is reduced, while the opposite is true apart for UTX stock. Furthermore, we can see that, apart for $\widehat{H}$ for the MRK stock, all the other estimates fall outside of the confidence intervals of the same parameters' estimates reported in Table \ref{tab_test}. This means that the estimated coefficients reported in Table \ref{tab_test} are statistically different (at $5\%$ significance level) from the ones estimated without the jumps. Results make clear that in case of abrupt jumps or spikes, the estimates of the scaling exponent parameters can be severely impacted.

\section{Practical application of scaling and multiscaling to VaR}\label{sec_var}
In this section, we show that by using a simple VaR configuration without a scaling or multiscaling consideration might bias the VaR estimation at higher aggregation scales. We use daily stocks data from the Dow Jones index to estimate the multiscaling spectrum and to carry out
 the multiscaling test described in Section \ref{sec_ea}.  After the procedure is concluded, we estimate the two most common VaR models, i.e. the Historical and Gaussian VaR at $1$ day. Successively, we use these estimates to compute the yearly VaR using the \textit{square root of time rule}.\footnote{Theoretically, this is only valid with the Gaussian formulation. Nevertheless, it is commonly used also for Historical VaR time aggregation.} We then compare them with the fractional VaR with proper scaling and highlight eventual biases. To conclude, we propose a methodology to compute a multiscaling consistent VaR.
\subsection{Value at Risk}
VaR is an easy and intuitive way to quantify risk for assets and portfolios. Let $VaR(\tau,1-\alpha)$ be the Value at Risk at frequency $\tau$ for a confidence level equal to $1-\alpha$ which satisfies
\begin{equation}
P(r_\tau(t)<VaR(\tau,1-\alpha))=\alpha
\end{equation}
where $r_\tau(t)$ are the log-returns at frequency $\tau$. Several methodologies are used to compute the VaR. Among all, we recall the Historical VaR (HVaR) and the Gaussian VaR(GVaR).\footnote{Besides the common knowledge of non-Gaussianity of stocks' log-returns, the Gaussian VaR is widely adopted both in academia and industry.} The former is a non-parametric approach that uses historical data to compute the VaR, while the latter assumes a Gaussian distribution of stock returns and applies the Gaussian formula for the percentiles computation to extract the VaR at a given confidence level. The issue faced in applied Finance is that the \textit{square root of time rule} works only under the assumption of \textit{iid} Gaussian returns. However, this technique is widely adopted regardless of its assumptions. 

\begin{figure}[h!]
	\begin{center}	
		\includegraphics[width=0.8\textwidth,height=0.3\textheight]{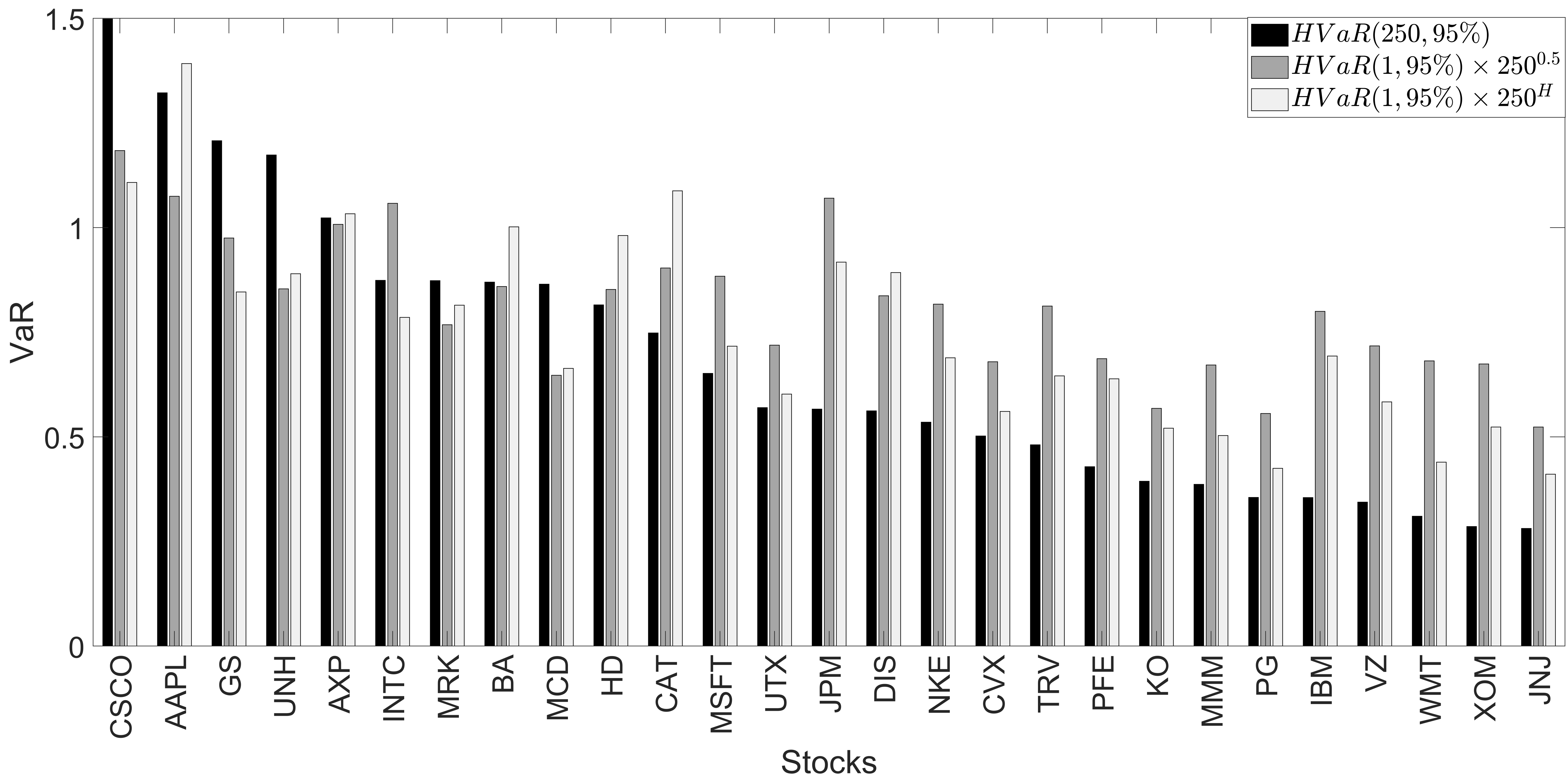} 

	\end{center}
	\caption{HVaR using annual data and using daily data rescaled by the factor $250^{0.5}$ and by $250^{\widehat{H}}$. Stocks are sorted by the magnitude of the annual Historical VaR.}
	\label{VaR}
\end{figure}

\begin{figure}[h!]
	\begin{center}	

		\includegraphics[width=0.8\textwidth,height=0.3\textheight]{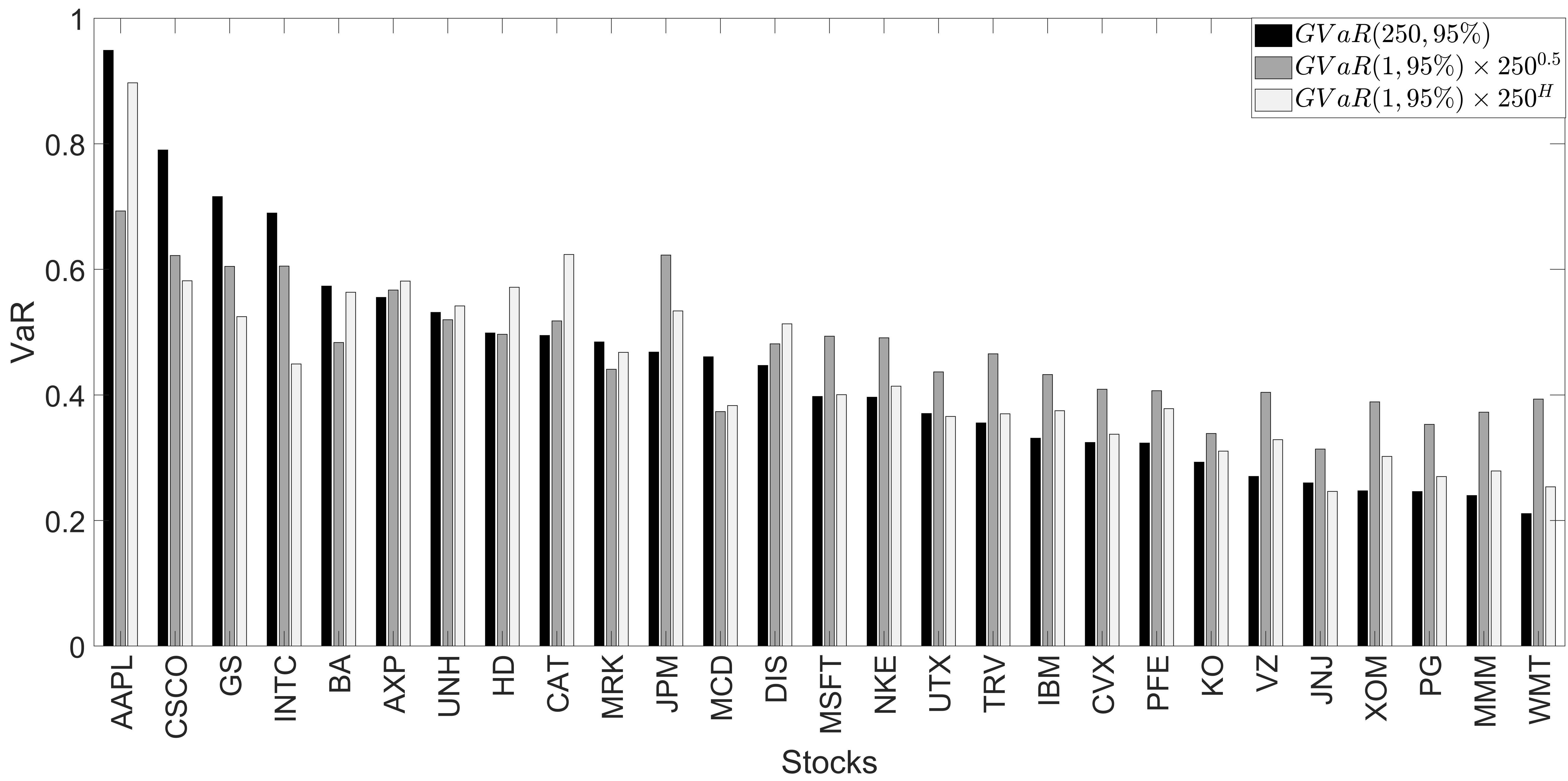} 
	\end{center}
	\caption{GVaR using annual data and using daily data rescaled by the factor $250^{0.5}$ and by $250^{\widehat{H}}$. Stocks are sorted by the magnitude of the annual Gaussian VaR.}
	\label{VaR2}
\end{figure}

In our analysis, VaR is computed using the two aforementioned approaches at $\tau=1$ day, and $95\%$ confidence level ($Var(1,95\%)$). Annual VaR ($Var(250,95\%)$) is calculated with the scaling exponent equal to $0.5$, i.e. $\widehat{Var}(250,95\%)=Var(1,95\%)\times 250^{0.5}$ and to the estimated $H$, i.e. $\widehat{Var}_H(250,95\%)=Var(1,95\%)\times250^{\widehat{H}}$.\footnote{We will discuss the multiscaling case later in the paper.} We further estimate the true $Var(250,95\%)$ using annual returns ($\tau=250$ days) and compare them. Results are shown in Figures \ref{VaR} and \ref{VaR2}.
These figures show that when we compare the VaR calculated using the $H$ scaling time rule and the VaR with \textit{the square root of time rule}, the deviation from the true VaR is lower when the former approach is used.
In fact, for the VaR with $H$ scaling time, the bias with respect to the true VaR is considerably lower for most of the stocks in both the HVaR and GVaR settings. This is due to the fact that over the long-run, even a small divergence from the assumption of scaling exponent equal to $0.5$ can have a substantial impact. 

\begin{figure}[h!]
	\begin{center}	
		\includegraphics[scale=0.15]{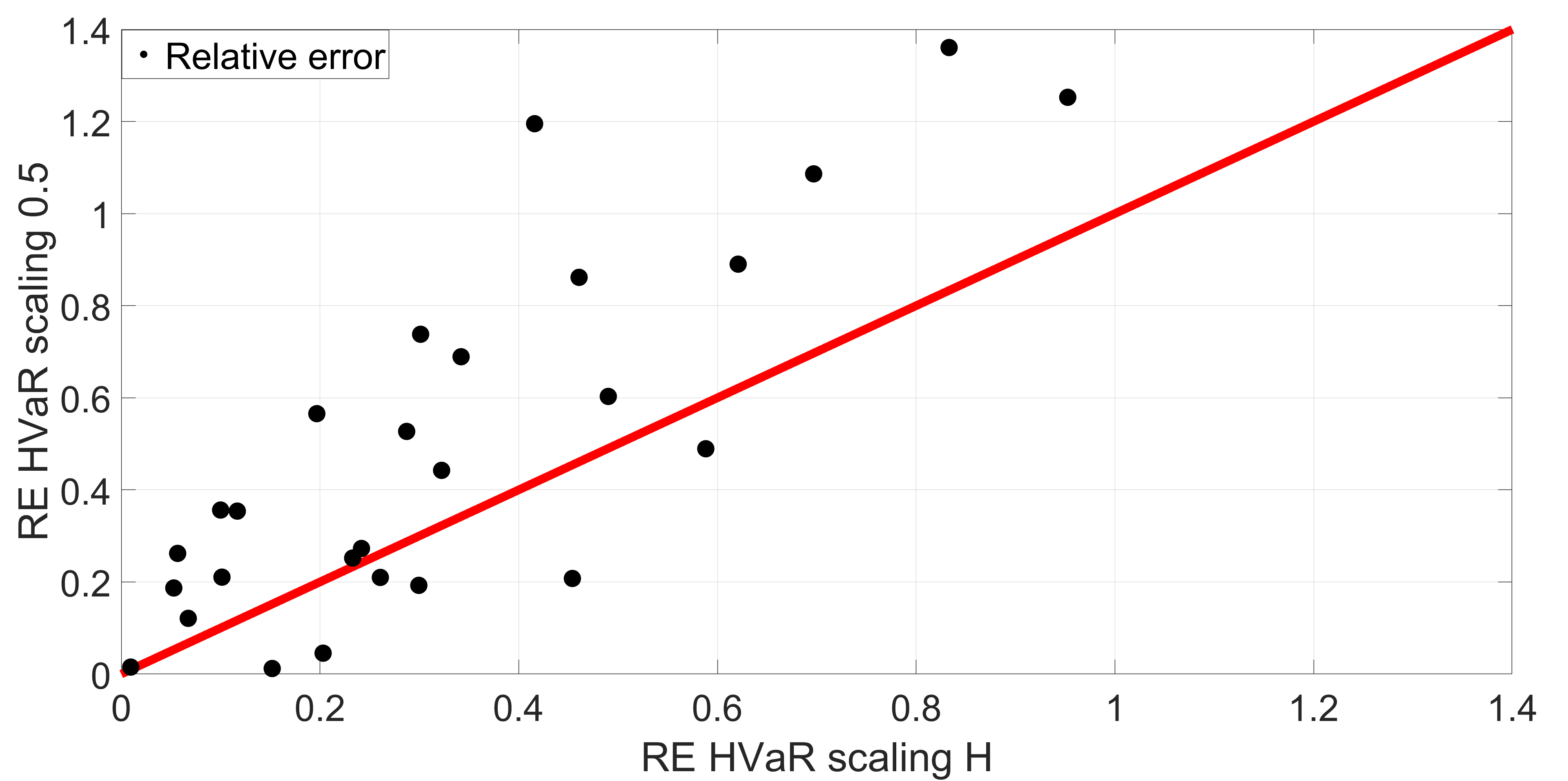} 
		\includegraphics[scale=0.15]{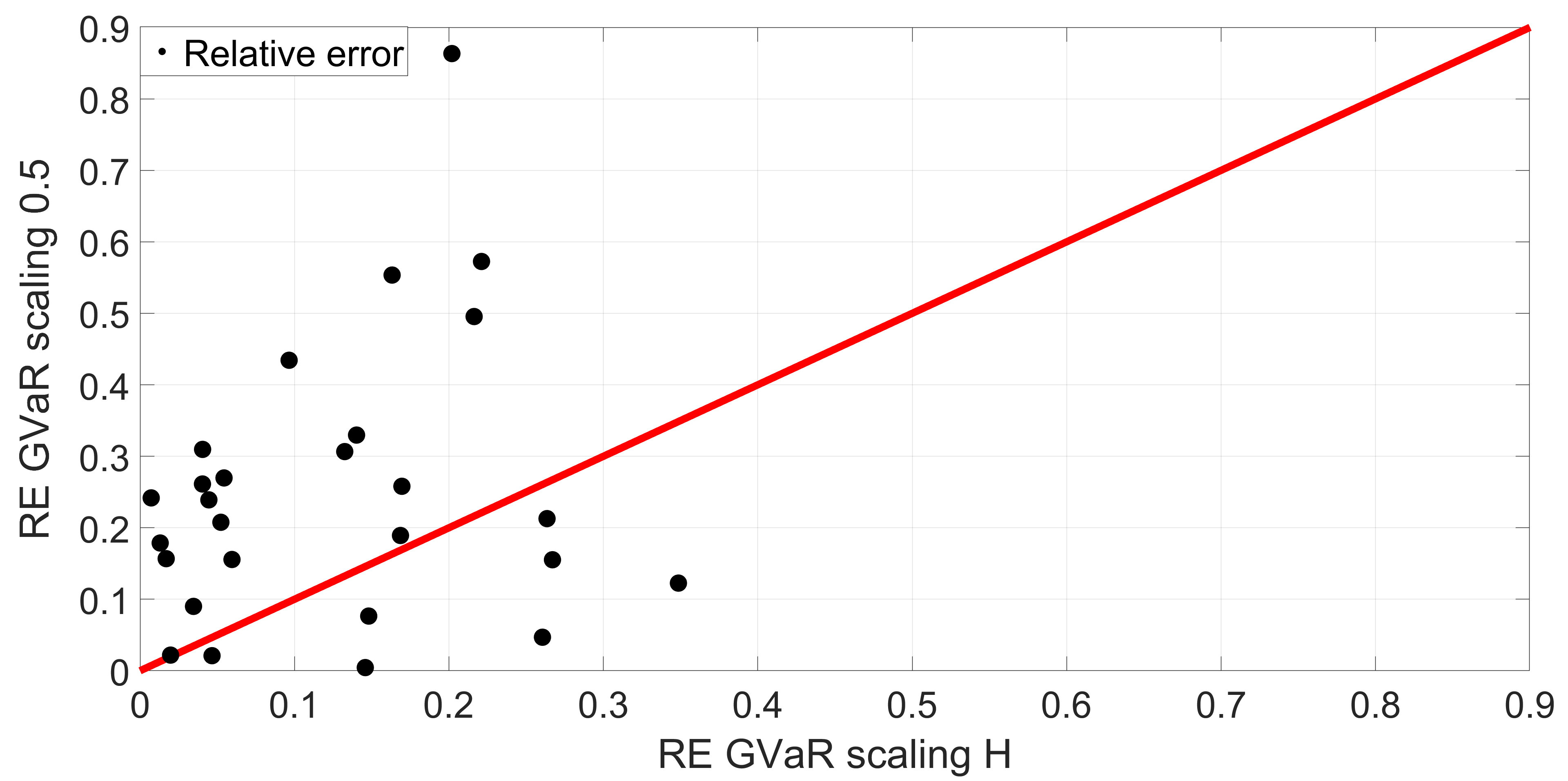} 
	\end{center}
	\caption{Relative error between the true VaR calculated using annual data, HVaR and GVaR computed using daily data scaled by the factor 0.5 and by the estimated factor $H$. The red line is the 45 degrees reference line.}
	\label{EVaR}
\end{figure}

To conclude the analysis with a quantitative assessment of the performance of the different methodologies, we also report the relative error (RE), i.e.
 \[RE=\frac{|Var(250,95\%)-Var(1,95\%)\times250^{K}|}{|Var(250,95\%)|},\]
  where $K$ is equal to $0.5$ for the VaR computed with the square root of time rule or equal to the estimated $H$, $\widehat{H}$, as reported in Table \ref{tab_test}. This helps to identify the magnitude of the deviation from the true VaR and to compare the two scaling approaches. Figure \ref{EVaR} shows the results.
As the figure shows, using the correct scaling results in a smaller relative error. This confirms that the choice of a proper scaling exponent should not be neglected by the financial community, considering that its estimation and testing are relatively simple. 
\subsubsection{Multiscaling consistent VaR}
\begin{figure}[h!]
	\begin{center}	
		
		\includegraphics[width=0.8\textwidth,height=0.3\textheight]{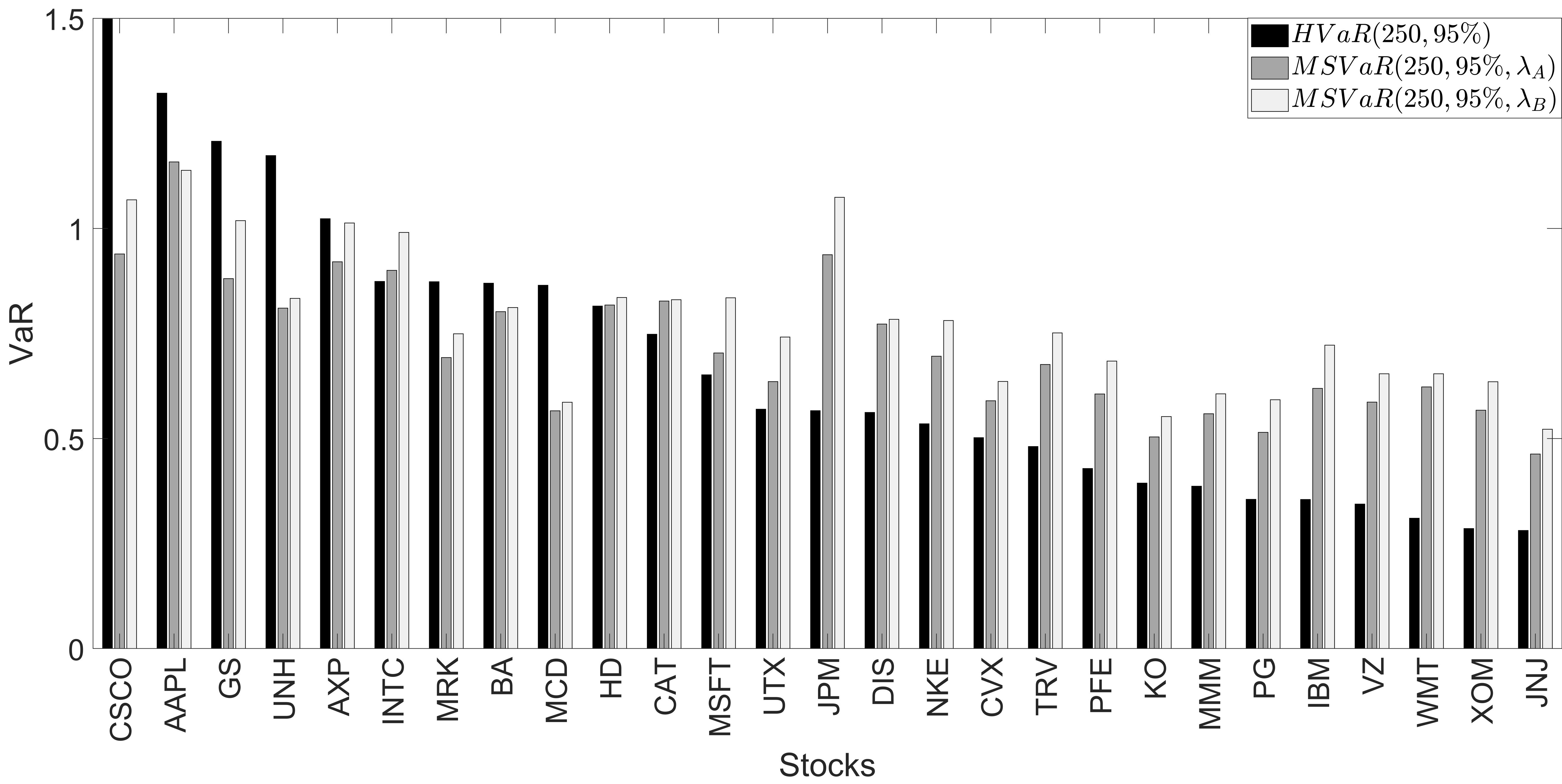} 
	\end{center}
	\caption{Annual Historical VaR (HVaR) and Multiscaling VaR (MSVaR). Confidence level $95\%$. Stocks are sorted according to the magnitude of the Historical VaR.}
	\label{VaR3}
\end{figure}

In the previous subsection, we showed that using the correct scaling contributes to reduce the computation error for VaR at smaller frequencies. However, as explained in Section \ref{mt}, all time series analyzed are strongly multiscaling. To deal with such situations, we discuss a possible solution. While VaR is related to the log-returns, multiscaling is a property of the moments of the log-returns. For this reason, there is not a straightforward formula to compute VaR which takes into account multiscaling. An exception to this is the Multifractal VaR proposed in \citep{lee2016multifractal}, where the author introduces a VaR consistent with the multifractality of financial time series using the Multifractal Model of Asset Returns (MMAR). In a previous paper \citep{batten2014}, a similar analysis is performed but it relies on the MMAR Monte Carlo simulations and it computes the VaR on the simulated time series. 
The Monte Carlo approach has the advantage of letting the researcher use the model that best depicts the data. In fact, one can calibrate the MRW or the MMAR and generate a large number of sample paths which can be used to compute VaR. In the case of moderate multiscaling, the difference can be low but for multiscaling processes with a $|B|>0.05$, neglecting such a feature can strongly distort the VaR. In this work, we use the MRW to simulate $250$ trading days (i.e. $1$ year) of log-returns and compute the VaR of the simulated paths. For this purpose, three parameters need to be estimated: the variance $\sigma^2$, the autocorrelation scale parameter $L$ and the intermittency parameter $\lambda$. The variance can be estimated from the log-returns time series as $\sigma^2=Var(r_{\tau}(t))$ with $\tau=1$, the parameter $L$ is set to be equal to $\tau^*$, while the intermittency parameter $\lambda$ can be extracted by equating the estimated coefficients of Equation \ref{mult_proxy2} to the parameters in Equation \ref{scaling} and getting two (possibly different) estimates of $\lambda$, i.e. $\lambda_A$ and $\lambda_B$.\footnote{If the data generating process is not a fully MRW, the estimation of $\lambda$ by using $A$ or $B$ can differ substantially. In our case, for most of the stocks analyzed the intermittency parameter computed with the two estimated coefficients, i.e. $\lambda_A$ and $\lambda_B$, are very similar.} For each stock, we estimate the three parameters, $\sigma^2$, $L$ and $\lambda$. Hence, we generate $100000$ independent paths of daily returns for a year (i.e. 250 days). Finally, we quantify the VaR, which we name Multiscaling VaR (MSVaR), by computing the $95\%$ percentile on these simulations. Results are depicted in Figure \ref{VaR3}.

It is possible to appreciate that the MSVaR computed on the simulated paths has a comparable size to the Historical VaR computed on annual data. It is also important to note that the values predicted using $\lambda_A$ and $\lambda_B$ are very similar, suggesting that the stocks log-returns can be adequately approximated by the MRW model. Nevertheless, we remark on the importance of the full multiscaling estimation and testing procedure which lead to the MSVaR. In fact, if the previous analysis is bypassed the estimated risk metrics can be severely biased and inconsistent.

\section{Conclusions}\label{sec_c}
In this paper, we propose a step-by-step procedure to robustly estimate and test multiscaling in financial time series. By rewriting the structure function in a convenient way we perform  multiple tests on the scaling spectrum and assess the statistical significance of multiscaling, discriminating between weak and strong multiscaling. We have shown the effect of anomalies in financial time series and studied the impact on the estimated scaling exponents. Moreover, we have shown how the use of proper scaling can help to reduce the error in the VaR forecasting at a smaller frequency with respect to the commonly used \textit{square root of time rule}. Finally, we have proposed a Multiscaling consistent VaR using a Monte Carlo MRW simulation calibrated to the data and on which the VaR is then computed. Results are encouraging and confirm the goodness of the proposed methodological approach. Multiscaling is a stylized fact which can make a difference in the assessment of risk measures and in building quantitative models. It can be easily extrapolated from data and should not be overlooked by risk managers and authorities.

\section*{Acknowledgement}
 We would like to thank the anonymous referees who provided useful and detailed comments on a previous version of the manuscript. Their comments significantly improved the quality of this work. We want to thank the ESRC Network Plus project 'Rebuilding macroeconomics'. We are grateful to NVIDIA corporation for supporting our research in this area with the donation of a GPU. We thank Bloomberg for providing the data.
\bibliographystyle{tfcad}
\bibliography{StatsHurst_final_a_v2}
\end{document}